\documentclass[twocolumn]{aastex63}
\usepackage{amsmath,color,textcomp,url,graphicx}

\usepackage{subfigure}
\usepackage{savesym}
\savesymbol{tablenum}
\usepackage{siunitx}

\newcommand{\kms}{\hbox{km\,s$^{-1}$}}
\newcommand{\ms}{\hbox{m\,s$^{-1}$}}
\newcommand{\mjup}{$M_{\mathrm{Jup}}$}
\newcommand{\rjup}{$R_{\mathrm{Jup}}$}
\newcommand{\msol}{$M_{\odot}$}

\newcommand{\masyr}{$\mathrm{mas}\,\mathrm{yr}^{-1}$}

\submitjournal{PASP Tutorials}

\shorttitle{A Quick Guide to Nearby Young Associations}
\shortauthors{Gagn\'e et al.}

\begin{document}

\title{A QUICK GUIDE TO NEARBY YOUNG ASSOCIATIONS}

\author[0000-0002-2592-9612]{Jonathan Gagn\'e}
\affiliation{Plan\'etarium de Montr\'eal, Espace pour la Vie, 4801 av. Pierre-de Coubertin, Montr\'eal, Qu\'ebec, Canada}
\affiliation{Trottier Institute for Research on Exoplanets, Universit\'e de Montr\'eal, D\'epartement de Physique, C.P.~6128 Succ. Centre-ville, Montr\'eal, QC H3C~3J7, Canada}
\email{gagne@astro.umontreal.ca}

\begin{abstract}

Nearby associations of stars which are coeval are important benchmark laboratories because they provide robust measurements of stellar ages. The study of such coeval groups makes it possible to better understand star formation by studying the initial mass function, the binary fraction or the circumstellar disks of stars, to determine how the initially dense populations of young stars gradually disperse to form the field population, and to shed light on how the properties of stars, exoplanets and substellar objects evolve with distinct snapshots along their lifetime. The advent of large-scale missions such as Gaia is reshaping our understanding or stellar kinematics in the Solar neighborhood and beyond, and offers the opportunity to detect a large number of loose, coeval stellar associations for the first time, which evaded prior detection because of their low density or the faintness of their members. In parallel, advances in detection and characterization of exoplanets and substellar objects are starting to unveil the detailed properties of extrasolar atmospheres, as well as population-level distributions in fundamental exoplanet properties such as radii, masses, and orbital parameters. Accurate ages are still sparsely available to interpret the evolution of both exoplanets and substellar objects, and both fields are now ripe for detailed age investigations because we are starting to uncover ever-closer low-density associations that previously escaped detection, as well as exoplanets and ever lower-mass members of more distant open clusters and star-forming regions. In this paper, we review some recent advances in the identification and characterization of nearby associations, the methods by which stellar ages are measured, and some of the direct applications of the study of young associations such as the emergent field of isolated planetary-mass objects.

\end{abstract}


\section{INTRODUCTION}\label{sec:intro}

\begin{figure*}
    \centering
    \subfigure[Exoplanets mass-separation distribution of known young exoplanets]{\includegraphics[width=0.48\textwidth]{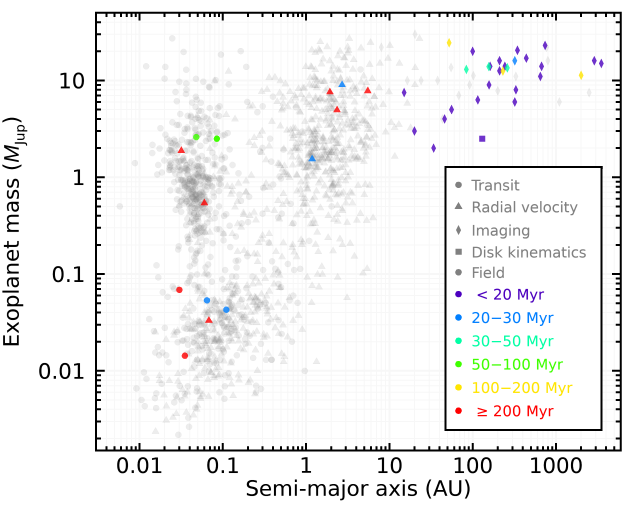}\label{fig:exo_mass}}
    \subfigure[Radius-period distribution of known young exoplanets]{\includegraphics[width=0.48\textwidth]{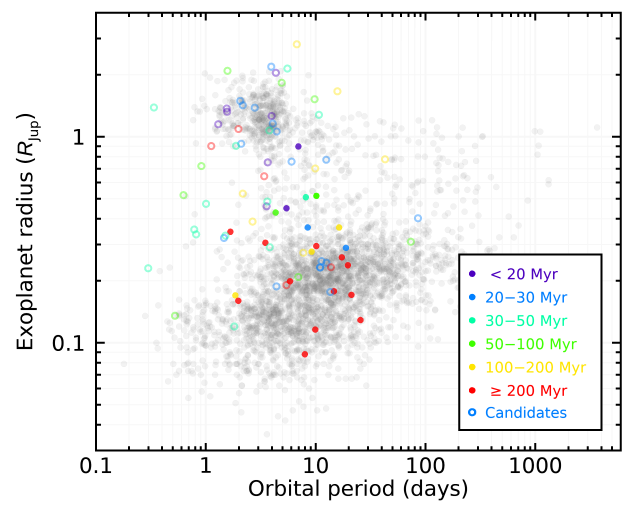}\label{fig:exo_rad}}
    \caption{Population properties of known exoplanets from the NASA exoplanet archive (gray symbols) compared with exoplanets in systems of known ages (colored symbols) based on their membership in a nearby young association. Most directly-imaged exoplanets were found by targeting known members of young moving groups, and the relatively sparse number of systems with known ages at short separations is a limiting factor to test theories such as atmospheric losses that drive the gap in detected exoplanets at $\approx 0.5$\,\rjup\ (the so-called `Neptunian desert'). See Section~\ref{sec:intro} for more details.}
    \label{fig:exo}
\end{figure*}

The measurement of stellar ages is notoriously difficult and often depend on various assumptions about the underlying physics of stars (see \citealp{2010ARAA..48..581S} for a review on the topic). The study of a collection of stars born together offers a unique opportunity to test age-dating methods against each other, using a wide range of stellar properties, and even to compare age-dating methods based on different types of astrophysical objects altogether. This is valuable not only to refine our understanding of stellar physics, but also to elucidate open questions about exoplanets and ultimately life in the Universe. For example, studying exoplanets around stars with accurately determined ages has the potential to answer some fundamental questions that are still open about the time scale of exoplanet formation \citep{2021ApJ...914..102C}, the orbital structures of exoplanets \citep{2019BAAS...51c.312C,2020DDA....5110103S} and how they evolve over time, the erosion of exoplanet atmospheres \citep{2021AJ....161..265D}, and how mantle degassing rates in rocky exoplanets can affect their atmospheres over time \citep{2022ApJ...930L...6U}. Answering these questions will be crucial to eventually identify systems where life as we know it could plausibly evolve, and to interpret potential biosignatures correctly.

The vast majority of stars are born in embedded clusters that typically contain more than a hundred members and form within the dense cores of giant molecular clouds of gas and dust \citep{2003ARAA..41...57L}. Within the first few $\approx$\,5\,Myr, most embedded clusters become gravitationally unbound because of the rapid dispersion of gas and dust resulting from the gas outflows of the forming protostars (see \citealp{2016ARAA..54..491B}), resulting in a collection of coeval stars with similar 3D velocities that are then subject to disruption and gradual mixing with the field population.

A minority of embedded clusters that form enough stars in the appropriate time scale can result in a fraction of their initial members forming gravitationally bound groups that emerge from the embedded phase as open clusters (e.g., \citealp{1983ApJ...274..698W,1984ApJ...285..141L}). Such clusters must first survive the tidal disruptions caused by their parent molecular clouds; additional tidal disruption events will then take place, caused by galactic tides and interstellar clouds and other open clusters encountered along their Galactic orbit. The duration for which open clusters can survive these various disruptions depends on their total masses, and can range from hundreds of millions to billions of years \citep{2002AA...389..871D,2005AA...441..117L,2021AA...645L...2A}.

Shortly after their emergence from their parent molecular cloud, these young stars display a narrow and structured velocity distribution because they formed in distinct embedded clusters that have not yet been disrupted even for the majority that became gravitationally unbound. This results in large complexes of young stars that still include short-lived O- and B-type stars. These ensembles were first recognized by the fact that OB stars are not randomly distributed over the sky, which led to their designations as `OB associations' (see e.g. \citealp{1995AstL...21...10M,1999AJ....117..354D} and references therein).

Because such OB associations include more than a single star-forming event, they can display significant age spreads. For example, \cite{2016MNRAS.461..794P} measured an age spread of $\approx$\,7\,Myr for the $\approx$\,15\,Myr-old Lower Centaurus-Crux subregion of the nearby Scorpius-Centaurus OB association \citep{1946PGro...52....1B}. OB associations therefore contain a number of overlapping, gravitationally unbound coeval populations with a relatively narrow range of ages and 3D velocities, and in most cases one or more bound open clusters \citep{1985ApJ...294..523E}. For example, the $\approx 16$\,Myr-old Upper Centaurus Lupus (UCL) region \citep{1964ARAA...2..213B} of the Scorpius-Centaurus OB association hosts two open clusters: UPK~640 \citep{2020AA...633A..99C} and UPK~606 \citep{2019JKAS...52..145S}.

Given a few dozens of millions of years, these distinct populations eventually drift apart, and those gravitationally unbound expand to become loose, low-density groups of coeval stars. Those can be significantly harder to recognize because their low densities are overwhelmed by those of the field population, and therefore accurate velocities must be measured for a large number of stars in order to recognized these populations.

The advent of the Hipparcos mission allowed to begin uncovering such `moving groups' of stars in the immediate neighborhood of the Sun (e.g., see \citealp{1997Sci...277...67K,2001ApJ...562L..87Z,2000ApJ...535..959Z,2004ApJ...613L..65Z} or \citealp{2004ARAA..42..685Z,2008hsf2.book..757T} for relevant reviews on the topic), but it is only with the Gaia mission \citep{2016AA...595A...1G} that it became possible possible to catalog loose associations at distances much beyond 100\,pc (e.g., see \citealp{2017AJ....153..257O,2019AJ....158..122K,2019ApJ...877...12T,2021AA...645A..84M}) or with particularly low densities and intermediate ages in the immediate solar neighborhood \citep{2022ApJ...939...94M,2023ApJ...945..119G}. The rich data provided by Gaia also made it possible to unveil the full complexity of the velocity distribution and sub-populations in young OB associations as well as younger star-forming regions (see e.g. \citealp{2021ApJ...917...23K,2021ApJS..254...20L,2022ApJ...941..143K,2022ApJ...941...49K,2023ApJ...954..134K}).

Other ensembles of stars can display structured over-densities in velocity space, yet do not appear to consist of coeval stars born in a single star-formation event. These are often designated as `streams'. One such example is the Hercules stream, which comprises relatively old stars with heterogeneous ages \citep{2020MNRAS.499.2416A}.

To complicate this picture further, other loose, co-moving and coeval structures were recently identified in the Gaia era as constituting tidal disruption tails around open clusters (e.g., \citealp{2019ApJ...877...12T,2019AA...621L...2R,2019AA...621L...3M,2019AA...627A...4R,2022MNRAS.517.3525B,2021AA...647A.137J}), providing a picture of the disruption process of open clusters. A large number of such tidal disruption tails remain to be fully characterized, and are being actively found around a large number of open clusters in the vicinity of the Sun.  The identification of a large 90\,pc-long population of young and co-moving stars in the Orion star-forming region was also made possible with Gaia data, and suggest that some of these structures may form along a wide molecular cloud filament rather than in an embedded cluster. The term `relic filament' was proposed to designate these structures \citep{2019MNRAS.489.4418J}.

For the purposes of this work, we use the term `stellar groups' to designate any ensemble of stars that is not necessarily coeval but is kinematically co-moving. The term `stellar streams' is reserved here to those ensembles that are explicitly non-coeval.

\cite{2003ARAA..41...57L} provided the following guidelines to define various groups of stars: `Open clusters' are groups of gravitationally bound coeval stars with at least 35 members (required to remain stable over a hundred million years); coeval groups with 6 or less stars are designated as `multiple systems'; `young associations' refer to groups with 7--34 members, or with a larger number of gravitationally unbound members. However, in practice it can sometimes be hard to immediately estimate the mass of an open cluster's complete population and estimate their dynamical stability. This led to the use of other empirical definitions which require that open clusters simply display a low dispersion in total proper motion, sky coordinates and parallax, a well-defined color-magnitude sequence which indicates coevality, as well as more than `a dozen' members \citep{2020AA...633A..99C}. The key parameter which can discriminate between open clusters and other young association is therefore the spatial density of their members, where the full population of open clusters is contained within approximately 15\,pc, whereas other moving groups can span hundreds of parsecs (e.g., see the Pisces-Eridanus group from \citealt{2019AA...622L..13M}).

We also use the term `moving group' to designate young associations, because this term has been widely used in the literature to discuss young associations within 100\,pc of the Sun (e.g., \citealp{2004ARAA..42..685Z} -- but note that `moving group' has also been used historically to designate OB associations). We use the term `star-forming regions' to designate groups of stars and gas clouds younger than $\approx$\,5\,Myr, and `OB associations' to designate groups where the gas and dust is mostly dissipated, but young enough that they retain a large number of overlapping loose associations (typically $\approx$\,6--30\,Myr).

We reserve the use of the term `tidal tail' to designate the loose structure around an open cluster or globular cluster that develops around it as its members are gradually disrupted (e.g., see \citealt{2003AJ....126.2385O}). The term `corona' was also used in the literature to designate groups of loose stars associated with an open cluster but located beyond their tidal radius (e.g., \citealt{1964SvA.....7..840A}), and was recently reintroduced by \cite{2021AA...645A..84M} to designate both tidal tails as well as loose groups around open clusters too young to have developed tidal structures (e.g., the $\approx$\,80\,Myr-old $\alpha$~Per cluster, \citealt{2022AA...664A..70G}). These looser structures correspond to unbound stars that probably formed in the vicinity of the open cluster, either in an embedded cluster or along a molecular cloud filament such as the relic filament identified by \cite{2019MNRAS.489.4418J}. They can therefore be expected to have a roughly similar age to the related open cluster, but they do not necessarily share the exact same age within a few Myr.

The study of open clusters and young stellar associations is particularly useful to understand how various astrophysical objects evolve over time. While open clusters and OB associations offer the opportunity to study large populations of coeval stars, only a few are located in the immediate vicinity of the Sun (the Coma~Ber, Hyades and Pleiades open clusters at 45--130\,pc, and the Scorpius-Centaurus OB association at $\approx$\,100--200\,pc\footnote{These ranges were obtained from a compilation of members with parallaxes at \url{http://mocadb.ca}.}). The study of other coeval populations of stars at closer distances is useful to investigate their faintest members such as substellar objects and exoplanets, but only a number of loose moving groups are located in our immediate neighborhood  within $\approx$\,70\,pc \citep{2004ARAA..42..685Z,2022ApJ...939...94M,2023ApJ...945..119G}, as well as an overlapping section of the more recently discovered Hyades tidal tails \citep{2019AA...621L...2R}.

Cataloging young stars near the Sun is a popular means to search for exoplanets with the method of direct imaging, whereas the contrast between the star's brightness and the exoplanet's intrinsic brightness are the fundamental limitation for detectability (see \citealp{2023ASPC..534..799C} for a detailed discussion on the topic). Young systems a particularly interesting in this scenario, because giant, gaseous exoplanets are still warm from their recent formation, and emit more infrared light, allowing us to detect them more easily (e.g., see \citealp{2004AA...425L..29C,2008Sci...322.1348M,2009AA...493.1149Z}). In fact, most of the directly imaged exoplanets discovered so far were found around young stars (see \citealt{2016SSRv..205..285M} and Figure~\ref{fig:exo}).

Refining our knowledge of stellar astrophysics, and in particular constraining evolutionary models at young ages and for the lowest-mass stars, is also important for the broader study of exoplanets. Indeed, exoplanets discoveries by the radial velocity \citep{2016ASSL..428....3H} and transit \citep{2009AA...506..287L,2015JATIS...1a4003R} methods yield high-accuracy measurements of the mass ratios and radius ratios of the exoplanet to its host star, respectively. The uncertainty of the exoplanet masses, radii and thus mass densities are still limited by the dominant measurement errors of the stellar radii and masses in most cases (e.g., see \citealt{2019ApJ...871...63M}). This bears significant consequences on our ability to determine the core composition of rocky exoplanets based on their bulk mass densities \citep{2011ApJ...738...59R}. The saying `Know thy star, know thy exoplanet' has recently become popular to express this need to better constrain stellar properties in the field of exoplanets.

\section{GALACTIC COORDINATES AND STELLAR KINEMATICS}\label{sec:coord}

One of the defining aspects of young associations is that they share a rough 3D spatial position in the Galaxy and revolve along similar Galactic orbits around the Milky Way, determined by the original orbit of the molecular cloud from which they have formed. Because the typical Galactic orbits of stars in the Solar neighborhood have periods of $\approx 240$\,Myr, their local trajectories on the time scales of decades can be approximated by straight three-dimensional lines. 

\subsection{Galactic Coordinates}
 
Re-constructing the three-dimensional position and velocity of a star from observational measurements requires a bit of calculation. The simplest case is reconstructing the 3D Galactic coordinates $XYZ$ starting from a star's sky coordinates (its right ascension $\alpha$ and declination $\delta$) and its distance from the Sun $d$, usually expressed in parsecs. Although calculating a Cartesian 3D position starting from two angles and a distance involves relatively simple geometry, we must be careful because sky coordinates and $XYZ$ coordinates are specified in different reference frames. Indeed, sky coordinate are defined in an equatorial reference frame where the grid of coordinates are aligned with the Earth's grid of latitudes and longitudes, whereas the $XYZ$ Galactic coordinates are aligned on the Galactic coordinate system, as the name suggests. Although different $XYZ$ Galactic coordinate systems are used in astronomy, the study of nearby stars usually rely on a right-handed coordinate system\footnote{This right-handed system is used by \cite{1987AJ.....93..864J} and has been commonly used since the Hipparcos mission, e.g. see \cite{1998MNRAS.298..387D,2007AA...474..653V,2012AA...538A..78L}.} centered on the current position of the Sun with $X$ pointing towards the Galactic center, $Y$ towards the global direction of rotation within the plane of the Milky Way, and $Z$ outwards and perpendicular from the plane in the Northern Galactic hemisphere.

Transforming a set of right ascension and declination and a distance to three-dimensional $XYZ$ Galactic coordinates therefore first requires performing a 3D rotation of the sky coordinates from the Equatorial ($\alpha$,$\delta$) to the Galactic ($l$,$b$) frame of reference, and then applying a simple geometric transformation \citep{1987AJ.....93..864J}:

\begin{align}
    \begin{bmatrix}
        X \\
        Y \\
        Z
    \end{bmatrix} = \begin{bmatrix}
        d \cos b\cos l \\
        d \cos b\sin l \\
        d \sin b
    \end{bmatrix},
\end{align}
where $d$ is the distance of the star.

\subsection{Trigonometric Distance}

A trigonometric distance is measured from a star's parallax motion with respect to distant background stars as the Earth progresses along its orbit around the Sun. Parallaxes are often measured in milliarcseconds (mas), and a trigonometric distance can then be obtained with the equation:

\begin{align}
    d {\,[\text{pc}]} = \frac{1000}{\varpi\,[\text{mas}]},
\end{align}
where $d$ is the trigonometric distance, and $\varpi$ is the parallax. Note that error propagation, and even determining the most likely distance, can be more complicated than the equation above suggests when the measurement errors are large, or the distances involved are much further than $\approx 100$\,pc (see \citealp{1973PASP...85..573L,1998MNRAS.294L..41O,2018AA...616A...9L} as well as \citealp{2021AJ....161..147B} and references therein); similar considerations apply to determining the most likely $XYZ$ Galactic coordinates of a star.

Measuring the kinematics of a star is usually done by comparing its sky coordinates at several epochs to measure the two components of its motion that are tangential to the plane of the sky (i.e., the proper motion), and then relying on spectroscopy to measure the radial component of its velocity (i.e., the radial velocity) using the Doppler effect.

\subsection{Proper Motion}

Proper motions are usually represented in the units of milli-arcseconds per year (\masyr), and can be calculated from the rate of change in sky positions between two reference epochs. The simplest example with two epochs (ignoring parallax motion) would be calculated using:

\begin{align}
    \mu_\alpha^\star &= \frac{1}{\num{3.6e6}}\cdot\frac{\alpha (t2) - \alpha (t1)}{t2-t1}\cdot\cos(\delta),\\
    \mu_\delta &= \frac{1}{\num{3.6e6}}\cdot\frac{\delta (t2) - \delta (t1)}{t2-t1},
\end{align}
where the proper motions are expressed in \masyr, ($\alpha(t)$,$\delta(t)$) is a set of sky coordinates (in degrees), measured at the epoch $t$ (in decimal years).

Note that the cosine of the declination only appears in the equation for the proper motion in the right ascension direction: this is a consequence of spherical geometry, where large changes in right ascension near the poles do not correspond to large changes in physical motion. This Jacobian cosine term is almost always implicitly included in measurements of proper motions, and any measurements that involve a change in right ascension, such as the measurement errors of a star's right ascension. The $^\star$ symbol is sometimes used to explicitly indicate that the cosine term was included, but it is often omitted whereas the cosine term is generally included regardless.

\subsection{Radial Velocity}

Radial velocities are usually measured by comparing a star's spectrum with a reference spectrum for which the radial velocity is known, models of stellar spectra, or reference wavelengths of known chemical species. The Doppler equation is then used to translate this relative change in wavelengths to a radial velocity, often expressed in the units of \kms:

\begin{align}
    v_{\rm rad} = c\cdot\num{1e-3} \ln\left(\frac{\lambda_{\rm observed}}{\lambda_{\rm vacuum}}\right),
\end{align}
where $v_{\rm rad}$ is the radial velocity in \kms, $c$ is the speed of light in \ms, and $\lambda$ is the wavelength (expressed in any units). This equation ignores relativistic effects (not important at the \kms\ scale), and positive values of $v_{\rm rad}$ indicate a shift to the red, and therefore an object moving away from the observer.

\subsection{Gravitational Redshift}

When measuring radial velocities, several phenomena must be corrected depending on the required accuracy \citep{2003AA...401.1185L}. Computing the Galactic space velocities $UVW$ requires knowing the radial velocity of a star around the solar system barycenter \citep{2014PASP..126..838W}, and the spectrum we collect at the telescope includes not only the motion of the star with respect to that barycenter, but also the motion of the Earth. The Earth's motion around the solar system barycenter ($\approx\pm 30$\,\kms) must therefore be corrected by using the sky position of the star and the Earth's orbit at the date of the telescope observations. Obtaining a radial velocity accuracy better than $\approx 1$\,\kms\ also requires correcting the impact of the star's gravitational redshift and the blueshift caused by convective motions at its surface.

The attempts to measure stellar gravitational redshift go back to the early days of Einstein's general relativity (see \citealp{1994AHES...47..143H} for a review) and have an important impact on the determinations of white dwarf radial velocities (e.g., see \citealp{1925PASP...37Q.158A,1954ApJ...120..316P,1979AJ.....84..650W}). The impact of gravitational redshift was also detected historically by comparing the average radial velocities of giants and dwarf stars in multiple systems and open clusters (see \citealp{1982JApA....3..383G,1997AA...322..460N,2011AA...526A.127P}).

\begin{figure*}[p]
    \centering
    \subfigure[Gravitational redshift]{\includegraphics[width=0.48\textwidth]{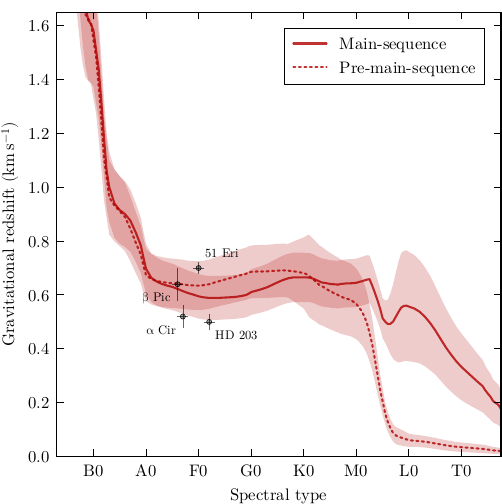}\label{fig:gshift}}
    \subfigure[Convective blueshift]{\includegraphics[width=0.48\textwidth]{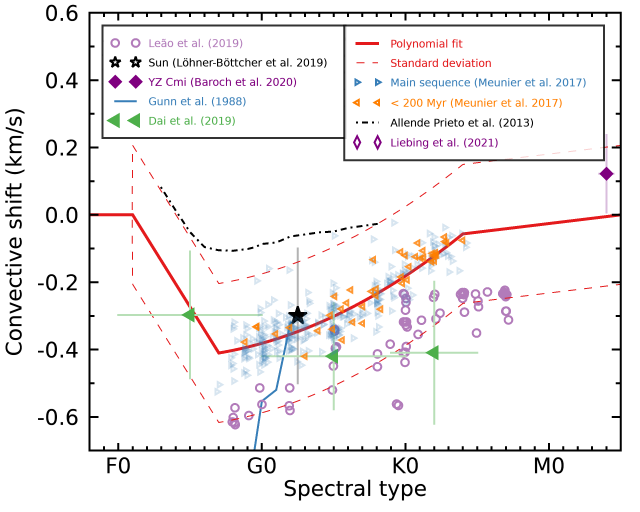}\label{fig:cshift}}
    \caption{Gravitational redshift (left) and convective blueshift (right) affecting the measured heliocentric radial velocities as a function of spectral types. The left panel is reproduced from \cite{2023ApJ...946....6C}. See Section~\ref{sec:act} for more details. Legend references: \cite{1988AJ.....96..198G,2013AA...550A.103A,2017AA...597A..52M,2019ApJ...871..119D,2019MNRAS.483.5026L,2019AA...624A..57L,2020AA...641A..69B,2021AA...654A.168L}.}
    \label{fig:rvshift}
\end{figure*}

\begin{figure*}[p]
    \centering
    \subfigure[Mass-spectral type sequences]{\includegraphics[width=0.48\textwidth]{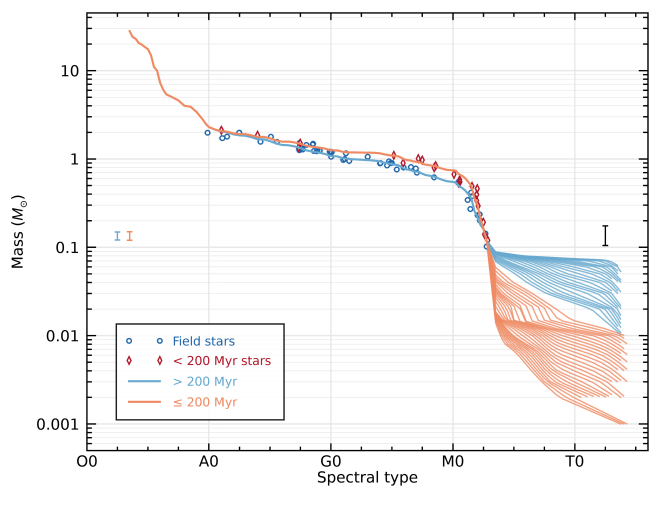}\label{fig:mass_seq}}
    \subfigure[Radius-spectral type sequences]{\includegraphics[width=0.48\textwidth]{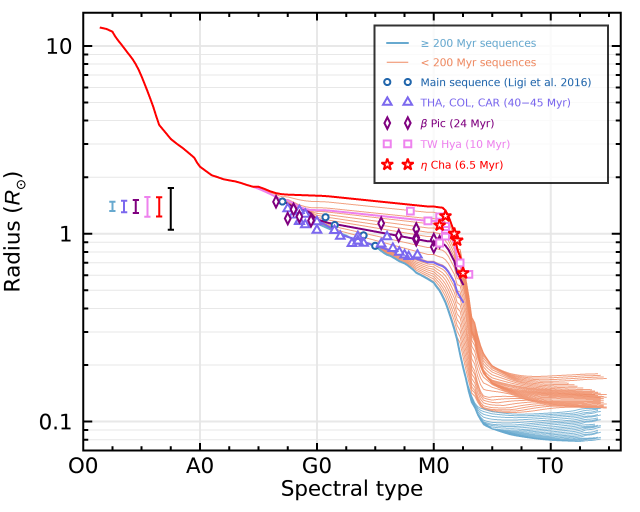}\label{fig:radii_seq}}
    \caption{Mass and radius sequences as a function of spectral type, based on \cite{2013ApJS..208....9P} data, supplemented with dynamical masses and interferometric radii of age-calibrated systems. See \cite{2023ApJ...946....6C} for details on the construction of the sequences, and Section~\ref{sec:coord} for more details.}
    \label{fig:mr_seq}
\end{figure*}

A star's gravitational redshift depends on its mass and radius and can be corrected with the following equation, which is an approximation arising from General Relativity in the context of weak gravitational fields:
\begin{align}
    \Delta v_{\rm rad; grav} = \frac{GM_{\rm star}}{cR_{\rm star}} = 0.636\,\hbox{km\,s$^{-1}$} \frac{M\,\hbox{[$M_\odot$]}}{R\,\hbox{[$R_\odot$]}},
\end{align}
where $G$ is the gravitational constant, and $\Delta v_{\rm rad; grav}$ is the gravitational redshift in \kms, which is defined as positive and must be subtracted to an observed radial velocity in order to correct it. The precision with which gravitational redshift can be corrected therefore depends on how precisely we can measure a star's mass and radius. It is generally hard to obtain gravitational redshift corrections with a precision better than $\approx$\,0.05\,\kms\ (see Figure~\ref{fig:rvshift}) for a single, isolated star because of uncertainties associated with its estimated mass and radius (e.g., see \citealp{2023ApJ...946....6C}). Note that this is especially challenging with young stars, where spectral type-mass and spectral type-radius relations built from main-sequence stars can be inappropriate even for approximate mass and radius determinations (see Figure~\ref{fig:mr_seq}).

\subsection{Convective blueshift}

Convective blueshift is caused by the fact that hot convective cells moving towards the star's surface are brighter and take up more fractional surface than the cooler interstices that are moving back towards the center of the star \citep{1981AA....96..345D,1982ApJ...255..200G,1990AA...228..203D}. As a consequence, stars that undergo significant convection will appear to move towards us (i.e., a Doppler blueshift) by as much as approximately 0.5\,\kms\ (e.g., see \citealp{2017AA...597A..52M}).

The convective blueshift is much harder to determine for an isolated star without an extensive study of its spectral line shapes with high-resolution spectroscopy (e.g., \citealp{2017AA...597A..52M,2019MNRAS.483.5026L,2021AA...654A.168L}). The extent of the impact of convective blueshift on the measured radial velocity depends on the spectral type of the star (see Figure~\ref{fig:rvshift}), but each line is affected to a different extent depending on its depth. Deeper spectral lines form closer to the stellar surface and are therefore less affected by convective blueshift because the velocity of convection is decreased near the surface \citep{2010ApJ...721..670G}. This effect combined with the fact that convective blueshift affects the shape of spectral lines (e.g., \citealp{1982ApJ...255..200G}) means that the impact of convective blueshift on the measured radial velocity is also a function of the instrumental resolving power, the numerical method and the wavelength range over which radial velocities are measured \citep{1987AA...172..200D,1999PASP..111.1132H}. This makes it challenging to correct convective blueshift in large radial velocity data sets, and can generally only be done with a statistical precision of $\approx 0.2$\,\kms\ (e.g., see \citealp{2023ApJ...946....6C}).

As a consequence of both gravitational redshift and convective blueshift, large radial velocity data sets are currently limited to precisions of $\approx 10$\,\ms\ for stars with a radiative surface (limited by the mass and radius estimates through gravitational redshift), and $\approx 200$\,\ms\ for those with a convective surface (limited by population-level estimates of convective blueshift). This is a stark contrast with the precisions often quoted for \emph{relative} changes in a star's radial velocity in exoplanet studies ($\approx 1$\,\ms; e.g., see \citealt{2020AA...636A..74T}).


\subsection{Galactic Space Velocity}

Once the sky position, distance, radial velocity and proper motion of a star is known, its three-dimensional $UVW$ Galactic space velocity can then be calculated with the following equation:

\begin{align}
    \begin{bmatrix}
        U \\
        V \\
        W
    \end{bmatrix} = \mathbf{R}(\alpha,\delta) \cdot\begin{bmatrix}
        v_{\rm rad}\\
        k\mu_\alpha/\varpi \\
        k\mu_\delta/\varpi
    \end{bmatrix},
\end{align}
where $UVW$ are expressed in \kms, proper motions in \masyr, the parallax in mas, $k = 4.74047$\,km\,yr\,s$^{-1}$, and $\mathbf{R}$ is a 3D matrix involving the sky coordinates of a star, described in \citet[Section 1.5.3]{hipb}.

\section{THE IDENTIFICATION OF STELLAR ASSOCIATIONS}\label{sec:ident}

Historically, nearby young associations of stars were found by identifying co-moving stars that are either massive and thus young, or have X-ray counterparts in the ROSAT catalog \citep{1999AA...349..389V}, indicating that they have enhanced stellar activity. These methods only allowed to uncover the tip of the iceberg in terms of the full population of young association members, and therefore only the youngest and most dense nearby moving groups were identified in this way (e.g., see \citealt{2004ARAA..42..685Z,2008hsf2.book..757T}).

Searches for additional members related to these sparse nearby young stars initially used methods such as the convergent point method \citep{1999MNRAS.306..381D}, utilizing the fact that stars with common $UVW$ Space velocities appear to be moving towards a fixed point on the sky when only their sky positions and proper motions are known.

\begin{figure*}
	\centering
	\includegraphics[width=0.95\textwidth]{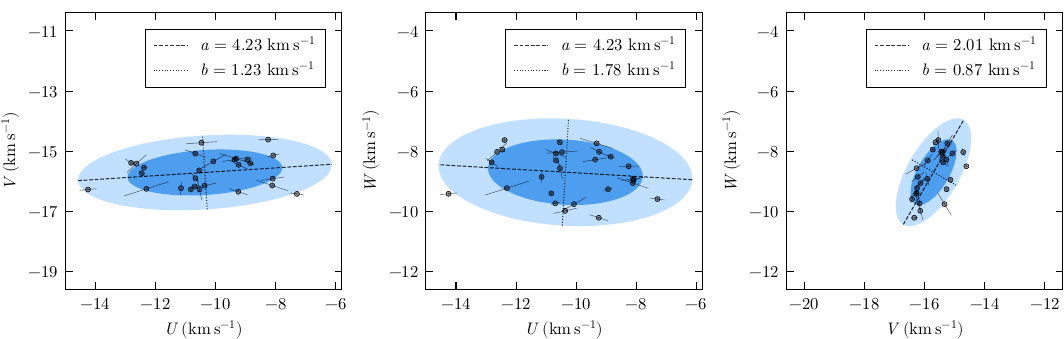}
	\caption{An example $UVW$ kinematic model of the $\beta$~Pictoris moving group (blue shaded ellipsoidal projections for the 67\% and 95\% probability contours) fitted to the individual members of the association (black circles with measurement errors). The typical uncertainties of stellar kinematics are currently dominated by measurement errors of the radial velocity since the advent of the Gaia mission, and those are oriented in the $UVW$ space in a way that depends on the star's sky position. Such multivariate Gaussian distributions in six dimensions ($XYZUVW$) are used in the model-selection tool BANYAN~$\Sigma$ \citep{2018ApJ...856...23G} used to determine the membership of a star in several nearby associations based on its measured kinematics. See Section~\ref{sec:ident} for more details. This figure is reproduced from \cite{2023ApJ...946....6C}.}
	\label{fig:banyan_model}
\end{figure*}

Methods based on Bayesian statistics and model selection were also gradually introduced (e.g., see \citealp{2013ApJ...762...88M,2014ApJ...783..121G,2018ApJ...856...23G}), which are especially useful to determine membership probabilities based on available kinematic observables in situations where radial velocities or parallaxes are missing, measurement errors are large or where stellar associations partially overlap. A limitation of these methods is that they rely on prior knowledge of the spatial distribution and $UVW$ space velocities of the associations being tested for membership (e.g., see Figure~\ref{fig:banyan_model}), and therefore they are not appropriate to identify new associations or spatial extensions to known associations.

Our understanding of stellar associations was significantly affected by gradual advances in the measurements of stellar kinematics on a large scale, most notably with the advent of the Gaia mission \citep{2016AA...595A...1G}. Their publication of more than 1 billion stellar parallaxes and more than 33 million radial velocities allowed calculating precise 3D Galactic coordinates $XYZ$ and Galactic space velocities $UVW$ for a larger number of stars than ever before.

\begin{figure*}[p]
    \centering
    \subfigure[$YZ$ projection of coronae near the Sun]{\includegraphics[width=0.48\textwidth]{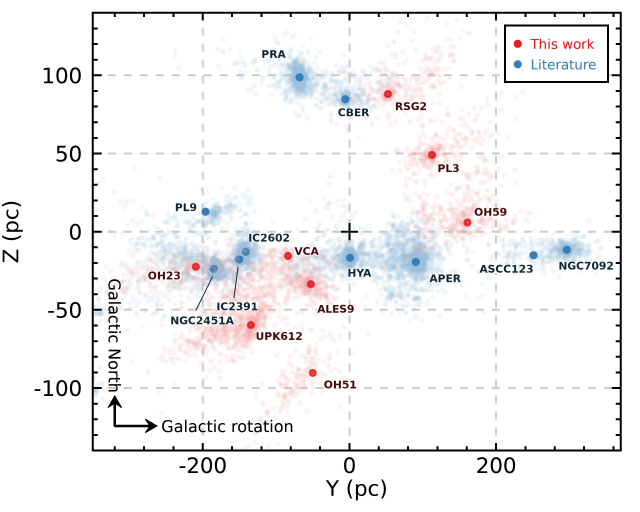}\label{fig:coronae_yz}}
    \subfigure[$XY$ projection of coronae near the Sun]{\includegraphics[width=0.48\textwidth]{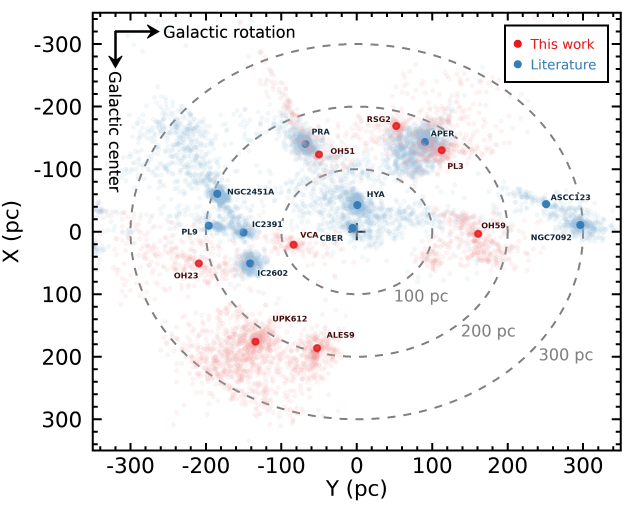}\label{fig:coronae_xy}}
    \caption{The spatial distribution of nearby open clusters with known coronae. See Section~\ref{sec:act} for more details. This figure is reproduced from \cite{2022ApJ...939...94M}.}
    \label{fig:coronae}
\end{figure*}

This rich data set made it possible to use modern machine-learning techniques such as DBSCAN \citep{1996kddm.conf..226E} and HDBSCAN \citep{campello2013density} to identify new co-moving stellar structures altogether. This not only allowed for the identification of most stellar members in nearby moving groups \citep{2018ApJ...860...43G,2018ApJ...862..138G}, but also for the discovery of entirely new structures named `coronae' that are related to known open clusters (e.g., \citealp{2021AA...645A..84M}, see Figure~\ref{fig:coronae} for an overview of coronae near the Sun), and a large number of entirely new groups of coeval stars (e.g., \citealp{2019AA...622L..13M,2019AJ....158..122K,2022ApJ...939...94M,2021AJ....161..171T,2022AJ....164...88B,2022AJ....164..115N}).

While some coronae found around more mature ($\geq 200$\,Myr) associations have been shown to correspond to tidal tails (e.g., \citealp{2019ApJ...877...12T,2019AA...621L...2R,2019AA...621L...3M,2019AA...627A...4R,2022MNRAS.517.3525B,2021AA...647A.137J}), others were found around associations such as the $\approx 80$\,Myr-old $\alpha$~Per cluster \citep{2022AA...664A..70G,2021AA...645A..84M}, much too young to have developed tidal tails.

Stars in these younger coronae may instead have formed along molecular cloud filaments \citep{2019MNRAS.489.4418J}, and may correspond to one formation mechanism of nearby associations which kinematics and ages are often similar to those of nearby open clusters (e.g., see \citealt{2021ApJ...915L..29G}).

\begin{figure}
	\centering
	\includegraphics[width=0.48\textwidth]{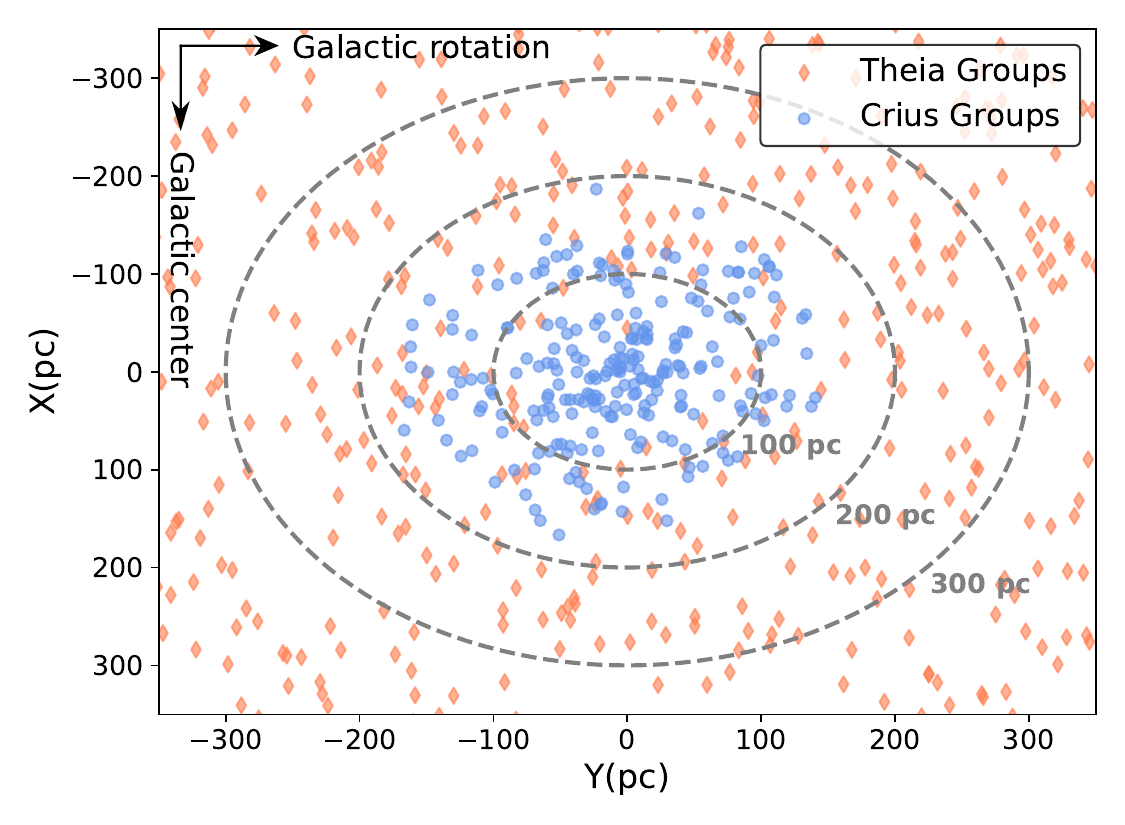}
	\caption{The projected $XY$ Galactic coordinates of groups recovered by HDBSCAN clustering methods based on only proper motions and 3D spatial position (orange diamonds, \citealt{2019AJ....158..122K}), and based on full $XYZUVW$ 6D kinematics (blue circles, \citealt{2022ApJ...939...94M}). The earlier case suffers from low recovery rates near the Sun because of geometric projection effects, whereas the latter case suffers from low recovery rates at further distances because of a lack in available high-quality heliocentric radial velocities required to build the 3D $UVW$ Galactic space velocities. See Section~\ref{sec:ident} for more details. This figure is reproduced from \cite{2022ApJ...939...94M}.}
	\label{fig:crius}
\end{figure}

One current limitation of clustering algorithms such as HDBSCAN is the fact they cannot account for measurement errors; this limits their applicability in that we must either search for over-densities in 6-dimensional $XYZUVW$ space with a much more limited data set (heliocentric radial velocities are now the bottleneck in measuring accurate $UVW$ space velocities for stars fainter than $G \approx 14$), or in 5-dimensional space ($XYZ$ Galactic coordinates and projected tangential velocities). While the latter method allows to search for co-moving structures at much larger distances (e.g., \citealt{2019AJ....158..122K}), geometric projection effects prevent them from successfully recovering most known co-moving structures within $\approx$\,70\,pc of the Sun (see \citealt{2022ApJ...939...94M} and Figure~\ref{fig:crius}).

This is especially a problem when one or more observable components of the 3D velocity are missing, most often the radial velocity. Because of their proximity and extended spatial size, the members of a nearby moving group can be located in vastly different directions as viewed from Earth, meaning that the missing radial velocity corresponds to different projected components of their similar 3D velocities. This fact greatly complicates the identification of over-densities in velocity space.

\begin{figure*}[p]
    \centering
    \subfigure[$XY$ distribution of young stars near the Sun]{\includegraphics[width=0.48\textwidth]{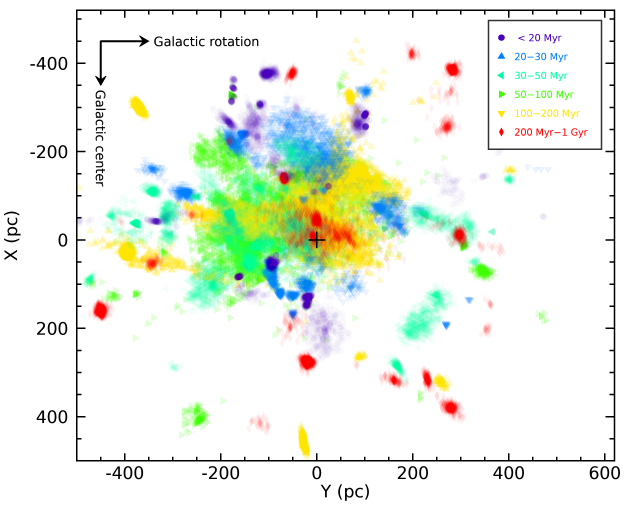}\label{fig:xy_asso}}
    \subfigure[$YZ$ distribution of young stars near the Sun]{\includegraphics[width=0.48\textwidth]{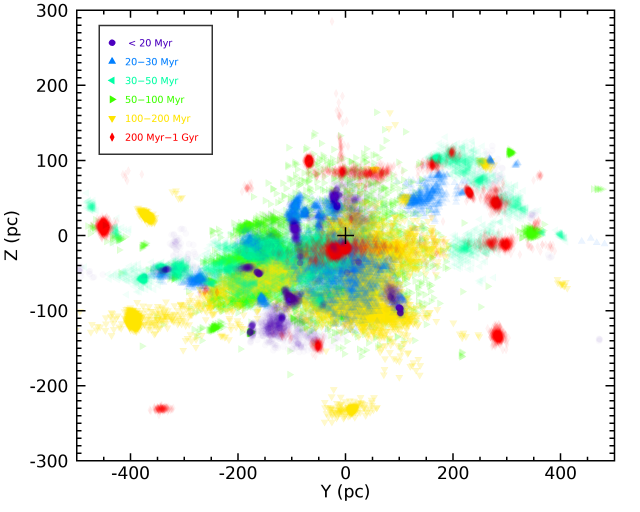}\label{fig:yz_asso}}
    \caption{Spatial distribution of nearby young stars, colored as a function of their ages. Older populations tend to be distributed more tightly because the looser ones have been mixed in with the rest of the field Galactic population. See Section~\ref{sec:ident} for more details.}
    \label{fig:xyz_asso}
\end{figure*}

\begin{figure*}
    \centering
    \includegraphics[width=0.9\textwidth]{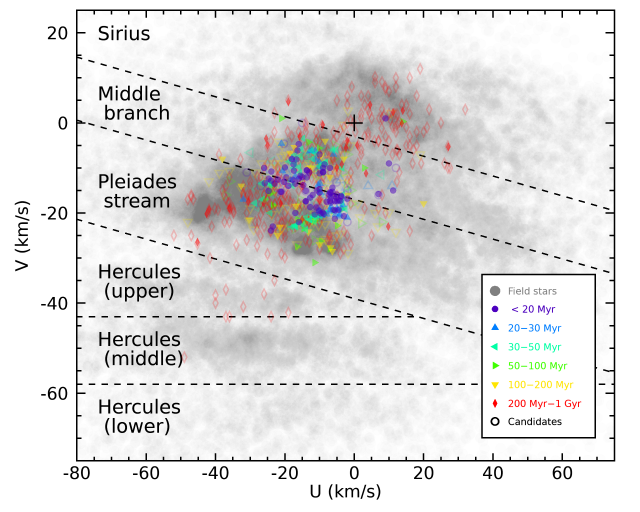}
    \caption{$UV$ space velocities of nearby field stars (gray) and young associations (colored symbols), with the branch structures of $UV$ plane delimited (see \citealt{2021AA...649A...6G} for a summary discussion, and \citealt{1958MNRAS.118..154E,1999MNRAS.308..731S,1998MNRAS.298..387D} for the specific delimitations in velocity space). See Section~\ref{sec:ident} for more details. Open symbols indicate candidate associations that have not yet been fully characterized in the literature. Symbol colors indicate different association ages.}
    \label{fig:uv_asso}
\end{figure*}

\begin{figure*}[p]
    \centering
    \subfigure[Ages and temperatures of stars in nearby associations]{\includegraphics[width=0.75\textwidth]{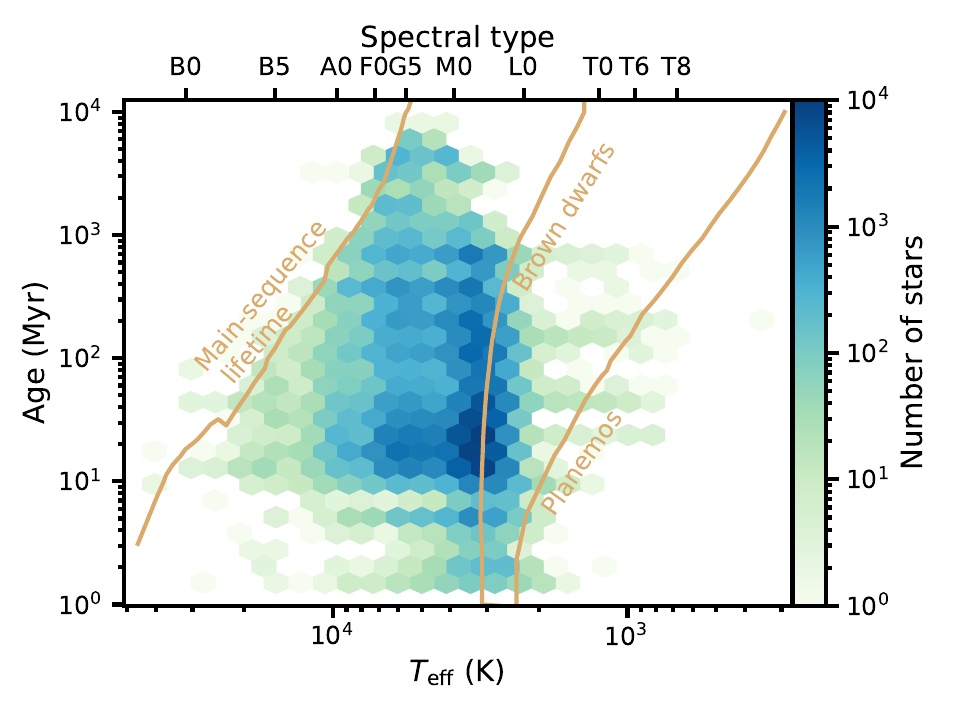}\label{fig:asso_teff_age}}
    \subfigure[Ages and distances of nearby young associations]{\includegraphics[width=0.75\textwidth]{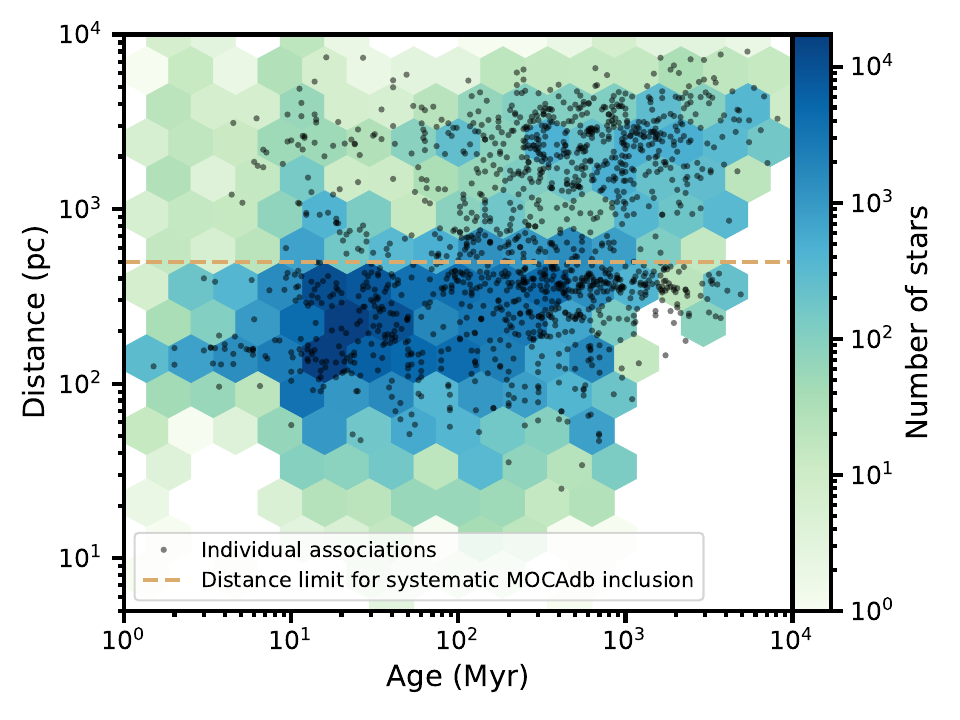}\label{fig:asso_age_dist}}
    \caption{The distribution of distances and ages of currently known young stars and associations near the Sun (complete to $\approx 500$\,pc only). Black circles indicate individual associations, and the colored hexagonal bins represent counts of individual stars. The stars and associations shown above are a compilation from the MOCA database of young associations (Gagn\'e et al., in preparation), compiling literature data from \citealt{2019AJ....158..122K,2022ApJ...939...94M,2021ApJ...917...23K,2020AA...633A..99C,2020AA...635A..45C,2004ApJ...613L..65Z,2021AA...645A..84M} among others. See Section~\ref{sec:ident} for more details.}
    \label{fig:asso_summary}
\end{figure*}

The Montreal Open Clusters and Associations (MOCA) database (Gagn\'e et al., in preparation) recently became available as a repository for extensive lists of members and candidate members in nearby young associations, along with their observed and inferred properties (e.g., spectral types, kinematics, photometry, rotation periods, activity indices, spectroscopic equivalent widths, masses, radii). The MOCA database is mostly complete to 500\,pc in terms of the included associations, but some clusters further away are also already included. The MOCA website\footnote{\url{http://mocadb.ca}} includes various tools to explore and download the data, including with basic MySQL queries, and the database can both be accessed with other database managers or directly through Python with the \texttt{pandas} dataframes\footnote{\url{https://github.com/jgagneastro/mocapy}}. An overview of the current map of young stars within 500\,pc of the Sun from the MOCA database is shown in Figures~\ref{fig:xyz_asso} and \ref{fig:uv_asso}. The ages, distances and temperatures of young stars and associations within 500\,pc of the Sun are also displayed in Figure~\ref{fig:asso_summary}.

\section{AGE ESTIMATION}\label{sec:age}

Measuring the ages of stars is one of the most famously difficult aspects of modern astrophysics \citep{2000ASPC..198..105M,2010ARAA..48..581S}. Various methods have been identified to successfully estimate the age of a star, but they are usually quite limited in their precision, and can most often only be applied for a range of stellar parameters. The use of a large coeval set of stars with varied properties can therefore provide an incredible advantage because not only the age-dating methods can be applied on a wider sample, but a range of different age-dating methods can be combined together to refine the age of the population as a whole. In this Section, I will briefly go over the current age-dating methods that have been successful in constraining the ages of nearby young associations.

\subsection{Isochrones}\label{sec:iso}

\begin{figure*}
	\centering
	\includegraphics[width=0.95\textwidth]{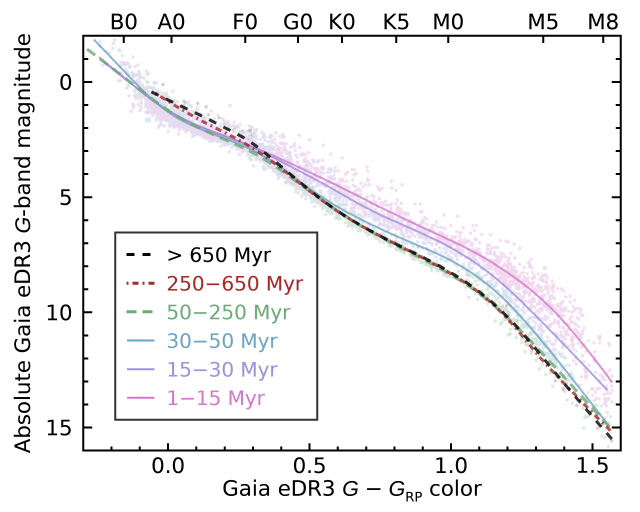}
	\caption{Gaia~DR3-based color-magnitude diagram of young stars with calibrated ages. Young low-mass stars appear over-luminous because their radii are inflated with respect to older stars, as they are still settling onto the main sequence. Conversely, short-lived massive stars inflate as they leave the main sequence, and therefore appear gradually brighter at older ages. See Section~\ref{sec:iso} for more details.}
	\label{fig:iso}
\end{figure*}

One of the most straightforward ways to estimate the age of a coeval population of stars is to investigate their distribution in a color-magnitude diagram such as the one shown in Figure~\ref{fig:iso} (e.g., see \citealp{2004AA...418..989N} and references therein). Stellar associations with gradually older ages tend to follow narrow sequences in such a figure: lower-mass stars are initially brighter at young ages because of their inflated radius, and they gradually become fainter as they slowly contract onto the main sequence. This happens much faster for Sun-like stars where they reach the main sequence by a few hundreds of millions of years and remain there for billions of years (e.g., \citealp{1997ApJ...482..420L,2016ApJ...823..102C}), limiting their use in this type of age diagnostic. For more massive stars, the trend reverses because they quickly reach the main sequence, deplete their nuclear fuel and their radii start inflating as they become red giants, causing them to appear brighter and cooler (thus redder) in a color-magnitude diagram (see \citealp{2022ARAA..60..455E} for a review on the topic). The shape and location of the color-magnitude diagram as a whole, or simply the turn-off point where all stars bluer (more massive) have already left the main sequence, can therefore be used as a powerful age-dating diagnostic when compared with either a reference association with a known age, or a model isochrone (e.g., \citealp{1993AJ....105.1420D}).

The extension of this method to gradually lower-mass stars has unveiled significant tension in the resulting ages based on model isochrones, compared with either the isochrone ages of higher-mass stars or other age-dating methods. This tension was resolved by accounting for strong magnetic fields in active, young low-mass stars which affect their interior structure and inflate their radii, making them appear brighter and thus younger in a color-magnitude diagram (e.g., see \citep{2012RAA....12....1P,2014prpl.conf..219S,2014ApJ...792...37M}). This, however, also introduced an additional free parameter (the magnetic field strength) in the fitting of model isochrones to young stellar populations.

Comparing the color-magnitude sequence of a young association to a set of other young association with well-calibrated ages based on a combination of age-dating methods can be a powerful tool to get away from such model systematics, and leverage the significant population of low-mass stars for age dating. Even in this scenario, however, one must be careful in correcting for interstellar extinction (mostly negligible within about 100\,pc of the Sun), and in considering that metallicity can also affect the shape and location of a color-magnitude diagram.

It is also interesting to note that associations younger than about 10\,Myr show a much more significant spread in their color-magnitude sequences, which may be a result of the wider spread in initial stellar rotation periods (e.g., \citealt{2015ApJ...807...58B}), and perhaps also in their initial radii.

\subsection{Lithium}\label{sec:li}

\begin{figure*}
	\centering
    \subfigure[Gradual lithium burning]{\includegraphics[width=0.48\textwidth]{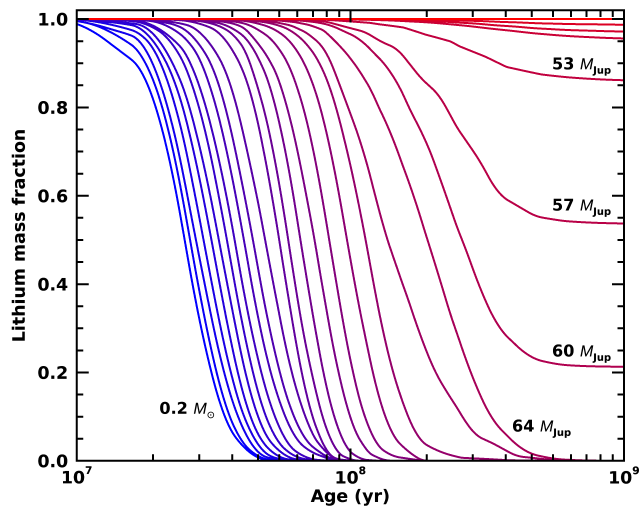}\label{fig:li_evol}}
    \subfigure[Lithium depletion boundary]{\includegraphics[width=0.48\textwidth]{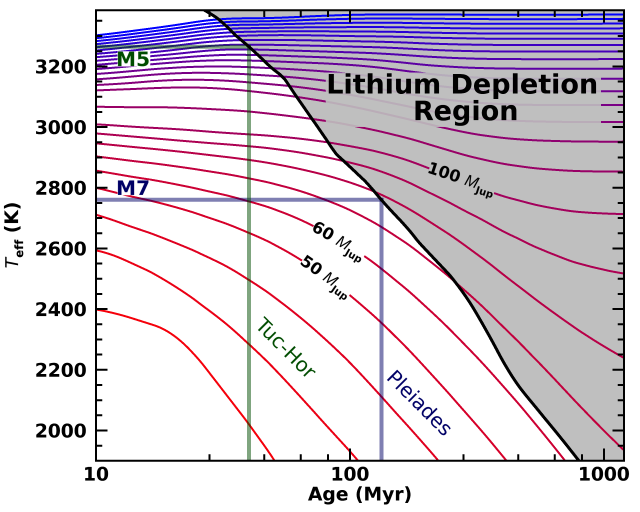}\label{fig:lel}}	
    \caption{The gradual lithium burning predicted by evolutionary models (left) and the resulting lithium depletion boundary as a function fo effective temperature and age (right). See Section~\ref{sec:li} for more details.}
	\label{fig:lels}
\end{figure*}

The burning of lithium in stars happens more efficiently and requires lower temperatures and pressures than the burning of hydrogen does, in part because no steps in the nuclear reactions require weak force interactions. As a consequence, a stellar core temperature of  \num{2e6}\,K is sufficient to initiate Li burning \citep{1996ApJ...469L..53R}. Stellar convection can act as a mixing mechanism, bringing Lithium from a stellar photosphere deeper in, where the temperature and pressure allows its burning, gradually lowering the lithium abundance in the photosphere over time. As a consequence, the measurement of lithium in a star's photosphere can provide precious indications on its age \citep{1991IAUS..145...85R,1991SSRv...57....1M}. The rate at which photospheric lithium is depleted, however, is a strong function of the stellar mass, because stars of different masses can have a significantly different internal structure (see Figure~\ref{fig:li_evol}.

In practice, the abundance of lithium is estimated using a measurement of the `equivalent width' of the \ion{Li}{1}~$\lambda$6708\,\AA\ spectral absorption line (e.g., \citealp{1993AJ....106.1059S}). The concept of an equivalent widths is an empirical tool to estimate the strength of a spectral line in a manner that does not depend strongly on the instrumental spectral resolving power, as long as the line is reasonably well resolved; a resolving power above 10,000 is required for best results in the case of lithium. The equivalent width is obtained by determining how wide in wavelength units a rectangle of uniform flux would be required to contain the same integrated spectral flux than the line is subtracting from the star's continuum. In practice, the continuum around cool stars in particular are often hard to measure because a large number of spectral lines are blended, and in these scenario a `pseudo continuum' is estimated by interpolation in the immediate vicinity of the measured line. 

\begin{figure}
	\centering
	\includegraphics[width=0.488\textwidth]{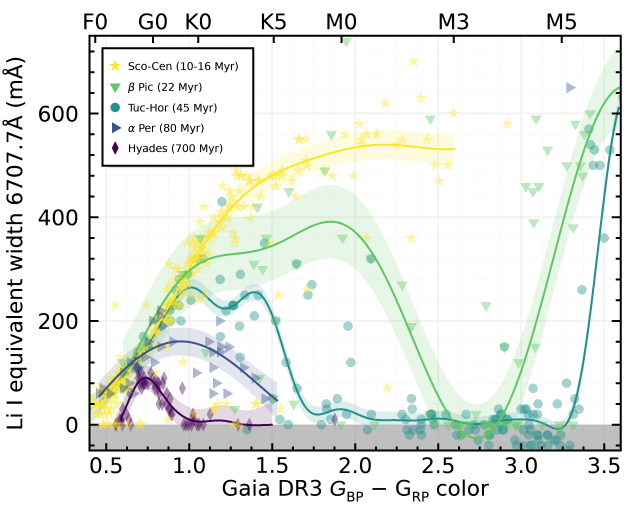}
	\caption{The observed distribution of lithium equivalent widths in the members of nearby age-calibrated associations. See Section~\ref{sec:li} for more details.}
	\label{fig:li_seq}
\end{figure}

Once the equivalent width of the \ion{Li}{1}~$\lambda$6708\,\AA\ line is measured, observers must rely on stellar models to estimate the actual lithium abundance in the photosphere of the star. For set of coeval stars more massive than about 0.3\,\msol, the sequence of lithium equivalent width as a function of temperature, color, or spectral type, can serve as a rough indicator of a population's age, as seen in Figure~\ref{fig:li_seq}.

For stars and brown dwarfs below 0.3\,\msol, however, lithium measurements become even more interesting. Because these objects are fully convective, photospheric lithium gets gradually depleted as soon as the core of the star or brown dwarf reaches the required temperature and pressure (Figure~\ref{fig:lels}). This causes a sharp transition in temperature above which lithium is suddenly depleted in a young coeval population of stars (see Figure~\ref{fig:li_seq}). Because brown dwarfs age as they cool, the location of this lithium depletion boundary remains a function of temperature even for populations old enough that only $\approx 60$\,\mjup\ objects still retain lithium \citep{1996ApJ...458..600B}. Locating this lithium depletion boundary for a population of coeval stars and brown dwarfs can therefore provide an accurate way of age-dating a population in the $20-200$\,Myr range with $3-8$\%\ uncertainties \citep{2004ApJ...604..272B}, in a way that was found to be relatively weakly dependent on the hypotheses required to build stellar and brown dwarf models (e.g., see \citealp{1996ApJ...458..600B} and Figure~\ref{fig:li_seq}, as well as \citealp{2022AA...664A..70G} and references therein for a detailed summary). Lithium depletion boundaries are harder to use for older populations, because it becomes harder to detect lithium in cool brown dwarf atmospheres.

\subsection{Gyrochronology}\label{sec:gyro}

The measurement of stellar rotation periods is a powerful tool to estimate the ages of stars, but does not work equally well for all types of stars (see \citealp{2009IAUS..258..345B,2009IAUS..258..375M} for reviews on the topic). When stars are born, the contracting gas spins gradually faster by conservation of its angular momentum, meaning that a star rotates gradually faster during its first few millions of years where contraction is still significant \citep{2010AA...520A..15M}. However, this initial contraction slows down as the star reaches the main sequence. At this point in their evolution, young stars typically show a wide range of different rotation periods.

\begin{figure}
	\centering
	\includegraphics[width=0.488\textwidth]{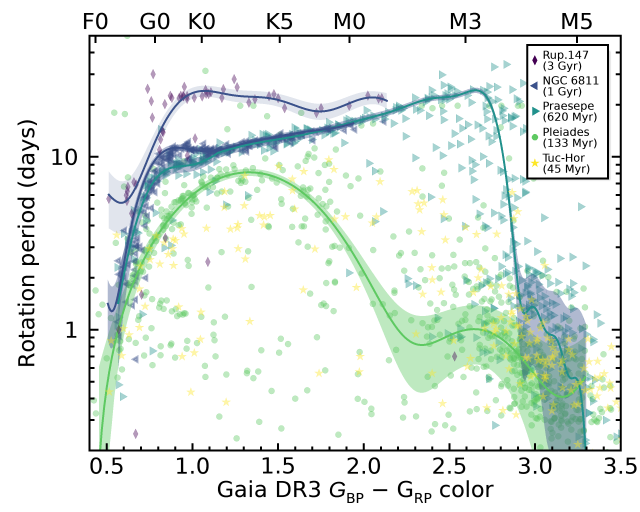}
	\caption{Rotation periods as a function of age for members of nearby associations with known ages. See Section~\ref{sec:gyro} for more details.}
	\label{fig:prot_seq}
\end{figure}

Stars with masses in the range $0.5-1.5$\,\msol\ (spectral types $\approx$ F0 to M3) have outer convective envelopes with an inner radiative core, and they initially feature a small-scale magnetic field driven by their convective layer (e.g., see \citealp{1969AN....291...49S}). Magnetic fields are able to drive `storms' where charged particles are ejected from the stellar surface at high velocities. These ejections interact with the magnetic field as they leave the surface of the star, stealing a small fraction of the star's angular momentum, a phenomenon called `magnetic breaking' (e.g., see \citealp{1991ApJ...376..204M}). Because the small-scale magnetic field is initially only coupled to the convective layer of the star, only this outer portion is slowed down. This period in a star's lifetime corresponds to the rapid-rotation sequence (sometimes called the `C' sequence) that can be observed in the younger associations such as the Pleiades, e.g., the Pleaides members in Figure~\ref{fig:prot_seq} with a rotation period of 1 day (F- and G-type stars) or less (K0--M3 stars).

After a period that ranges from a short time for early-F stars with very shallow convective zones, to 300\,Myr for early M dwarfs that feature deeper convective zones, the velocity shear at the interface between the two zones triggers a dynamo that pumps significant energy into the magnetic field of the star. This dynamo gives rise to a large-scale, bipolar magnetic field similar to that of the Sun that is coupled with the full star (e.g., see \citealp{1993ApJ...408..707P}). From this point on, the magnetic breaking starts slowing down the full mass of the star, and the efficiency of this breaking becomes more strongly dependent on stellar rotation. This strong dependency of breaking on rotation causes the stars with a wide range of initial rotation periods to converge onto a tight sequence for a coeval stellar population, often referred to as the  `I' sequence (e.g., K-type Pleaides members with rotation periods of 7-9 days in Figure~\ref{fig:prot_seq}). This `I' sequence then evolves to gradually slower rotation rates as the population ages. This makes it possible to use the sequence of stellar rotation as a function of spectral type or color to age-date a group of stars, a method called `gyrochronology' (e.g., see \citealp{1967ApJ...148..217W,1968MNRAS.138..359M,1988ApJ...333..236K,2003ApJ...586..464B,2013AA...556A..36G,2013PASJ...65...98S}).

Stars gradually migrate from the `C' to the `I' sequence as a population ages, and by the age of Praesepe ($\approx$\,620\,Myr; see Figure~\ref{fig:prot_seq}), most stars with outer convection zones (F0--M3) reached the narrower `I' sequence (stars later than M3 are fully convective). This makes gyrochronology a more powerful tool to assign accurate ages to stellar populations older than a few hundreds of millions of years, in contrast to other methods such as isochrones or lithium depletion.

Fortunately, rotation periods can be conveniently measured for stars in the range of masses where they are useful to constrain age, because these active stars tend to have numerous dark star spots on their surface, introducing a systematic modulation of their light curves as they complete a full rotation period \citep{2008ApJ...687.1264M}.

Stars with masses above 1.5\,\msol\ do not have a convective envelope necessary for magnetic breaking to slow stars down efficiently, and the outer layers of the atmospheres of stars much below 0.5\,\msol\ are too cool to be ionized, meaning they do not interact efficiently with the star's magnetic field. As a consequence, a large fraction of late M-type stars are fast rotators, very active magnetically, and retain the large spread in initial rotation periods they have gained during their stellar formation. The same observation can be made for brown dwarfs, which tend to rotate rapidly regardless of their ages \citep{2006ApJ...647.1405Z}.

\subsection{Stellar Activity}\label{sec:act}

Because stellar activity is a direct consequence of a star's magnetic field which is driven by stellar rotation, quantitative measurements of a star's activity can also be used to approximate its age \citep{1972ApJ...171..565S,1991ApJ...375..722S}. Several methods have been developed to characterize stellar activity, and empirical sequences have been built to estimate stellar ages from various activity indices (e.g., see \citealp{2008ApJ...687.1264M,2017MNRAS.471.1012B,2024ApJ...960...62E}).

Current evidence suggests that tiny and but extremely powerful `nanoflares', occurring regularly across the surface of the Sun, are responsible for heating up the Solar corona to extreme temperatures ($\approx 2$ million Kelvin, e.g., see \citealp{1965IAUS...22..390G,1974ApJ...190..447L,1988ApJ...330..474P,2015RSPTA.37340260C}). In such a scenario where an outer layer of gas is much hotter than the underlying regions, the ionized atoms in the hot layer contribute spectroscopic lines in emission. One such example is ionized hydrogen, which produces the prominent H$\alpha$ line often seen in emission at 656.281\,nm (e.g., \citealp{2005AA...431..329L}), as well as ionized Calcium Ca II which produces emission lines at 3968.47\,nm and 3933.664\,nm (the so-called Ca II H \& K lines; \citealt{1991ApJS...76..383D}), as well as a triplet at 849.8, 854.2, and 866.2\,nm (the so-called Ca II infrared triplet). Because magnetic fields are more intense in younger stars (due to their faster rotation leading to a more vigorous dynamo), these types of coronal emission lines are more common and are seen with more intensity when compared to stars of similar effective temperatures but older ages. Therefore, these emission lines are useful to estimate stellar ages, although they often vary in intensity over time. Measurements of the Ca II H \& K lines were shown to be present in emission in active stars as early as 1920 \citep{1920PASP...32..272H}, and were rapidly attributed to analogs of solar plages (bright regions of the solar corona located around its active regions \citealt{1913ApJ....38..292E}). The Ca II infrared triplet has become a more popular tool since the advent of infrared detectors, because it is located in a region of the spectrum not contaminated with telluric absorption from the Earth's atmospheres, and stars typically have a well-defined continuum in this region, making the measurement of an equivalent width easier \citep{2007AA...466.1089B}.

Another direct consequence of the extreme temperatures in active stellar coronae is emission of light in the UV and X-ray \citep{1996AA...305..284H}, which would not be expected to be significant given the effective temperature of these stars. The ROSAT space mission \citep{1982AdSpR...2d.241T} opened the doors to characterizing the X-ray luminosity of a large number of nearby, young stars and served as an extremely useful tool to identify the most active low-mass stars in the Solar neighborhood, although its limited sensitivity did not allow to observe the fainter M dwarfs at distances larger than $\approx 25$\,pc \citep{1998ApJ...504..461F}. The more recent eROSITA space mission \citep{2021AA...647A...1P} is more recently making it possible to characterize the X-ray emission of M dwarfs on a much larger scale, with about 25 times more sensitivity.

\begin{figure*}[p]
    \centering
    \subfigure[Ultraviolet]{\includegraphics[width=0.48\textwidth]{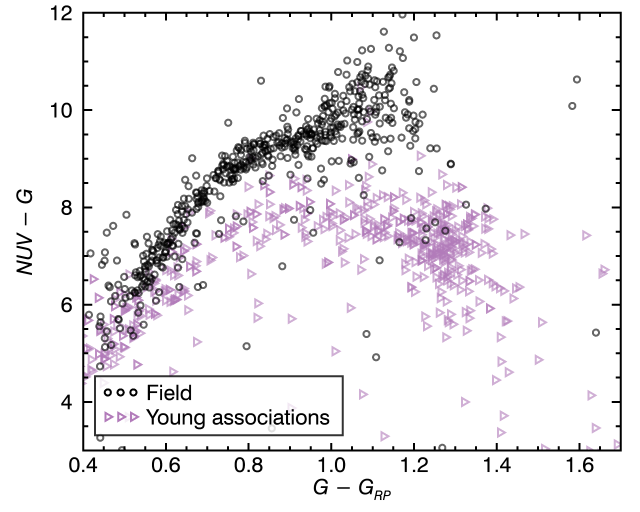}\label{fig:galex}}
    \subfigure[X-ray]{\includegraphics[width=0.48\textwidth]{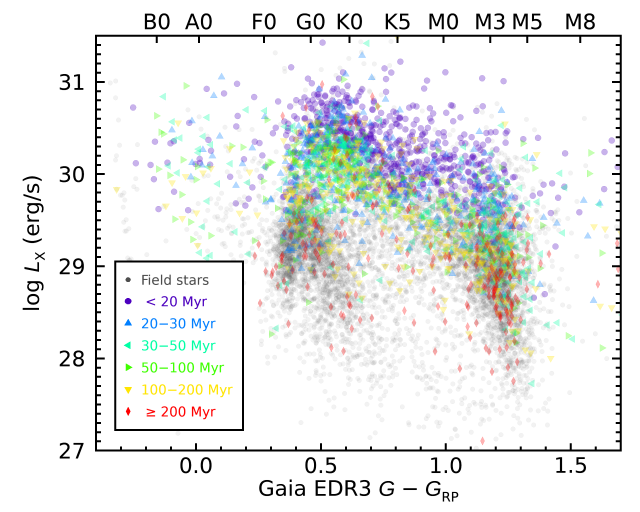}\label{fig:xray}}
    \caption{The distribution of near-UV colors (left) and X-ray luminosities (right) of young stars as a function of their Gaia colors. Stars younger than a few hundred Myr tend to have enhanced stellar activity due to their faster rotation rate, leading to brighter stellar plages, and thus bluer UV-to-visible colors and a brighter average X-ray flux. Their inflated radii also contributes to making their absolute X-ray luminosity significantly brighter than those of field stars at the same spectral type. See Section~\ref{sec:act} for more details.}
    \label{fig:uvx}
\end{figure*}

\begin{figure*}[p]
    \centering
    \subfigure[Ages and temperatures of stars in nearby associations]{\includegraphics[width=0.48\textwidth]{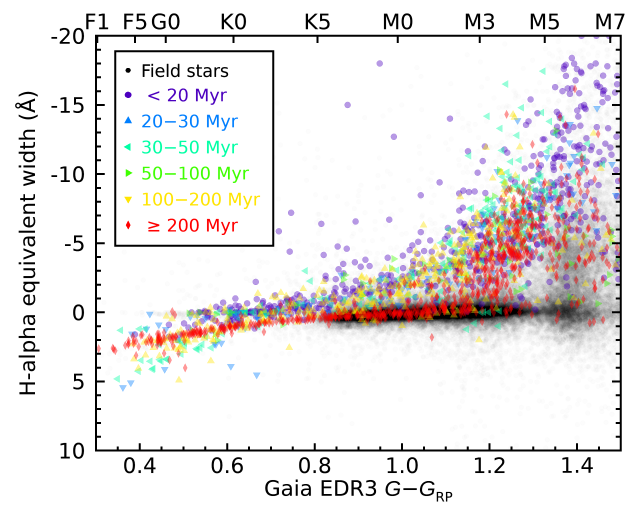}\label{fig:ha}}
    \subfigure[Ages and distances of nearby young associations]{\includegraphics[width=0.48\textwidth]{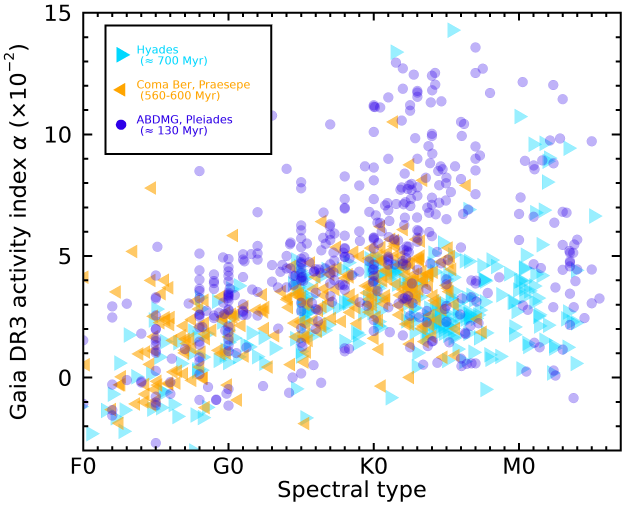}\label{fig:gaiaact}}
    \caption{The distribution of H$\alpha$ equivelent width measurements (left) and Gaia activity index based on the Ca~II infrared triplet for young stars (colored symbols) and field stars (black circles). Yet, again, the enhanced stellar activity is reflected in more pronounced H$\alpha$ and Ca~II emission lines produced by the hot stellar corona. See Section~\ref{sec:act} for more details.}
    \label{fig:act}
\end{figure*}

It is important to note that age measurements derived from stellar activity and gyrochronology are not independent of each other, so combining ages obtained from both methods does not guarantee that the resulting measurement errors are truly as small as one may expect when combining independent measurements.

\subsection{Tracebacks}\label{sec:iso}

Although the members of a young coeval association of stars share similar Galactic $UVW$ space velocities on average, they are born with an intrinsic spread in velocities that is a consequence of different initial gas velocities in the molecular cloud as well as the gravitational interactions of newborn stars with each other and the surrounding gas (see Figure~\ref{fig:banyan_model} for example). These random perturbations around the average molecular cloud's velocity will cause the stars to slowly spread apart spatially as they progress on their Galactic orbit. This offers the opportunity of determining the age of such a group of stars in a direct way that does not rely on astrophysical models a priori, and has therefore been the subject of extensive studies in the literature \citep{2002ApJ...575L..75O,2014AJ....148...70M,2022ApJ...941..143K,2023ApJ...946....6C,2020AA...642A.179M,2019MNRAS.489.3625C,2004ApJ...609..243O,2003ApJ...599..342S,2023MNRAS.519.3992Z}. By reconstructing the past trajectories of such a group of stars, we can determine the spatial size of the group at different moments in the past, and attempt to establish the moment at which the spatial size of the ensemble reaches a minimum (before the stars being spreading apart again, at past epochs that would correspond to moments before their birth). This method is commonly referred to as a kinematic traceback, and the moment in time where the group of stars is most compact corresponds to its traceback age. However, this method has often historically yielded unrealistically young ages when compared with other methods.

The main reasons for the disagreement between traceback ages and other methods have been the inaccuracy and the imprecision of kinematic measurements, sample contamination, and systematic errors introduced by unresolved multiple stars, as well as the systematic impacts of gravitational redshift and convective blueshift on the observed heliocentric radial velocities (see Section~\ref{sec:coord} and \citealt{2023ApJ...946....6C}). The advent of the Gaia mission dramatically improved the quality of tracebacks and renewed the interest in this method, partially closing the gap with other age-dating methods. The correct treatment of even small random measurement errors is also important, because the random motion that is otherwise introduced in the past trajectories can introduce systematic errors larger than the projected errors on the resulting traceback ages. This can be corrected either by numerical simulations or by using a forward model that is projected to the current epoch and compared with present-day kinematics directly (e.g., \citealt{2019MNRAS.489.3625C}). It is also plausible that traceback ages, even after systematic corrections, remain systematically younger than other age-dating methods by a $2-7$\,Myr, a period during which the physical size of the stellar association may not have changed significantly because of gravitational interaction between the stars and the interstellar gas before the latter was dispersed by radiation pressure of massive stars \citep{2021MNRAS.504..487K}. Recent work by \cite{2024NatAs...8..216M} estimated that this period preceding the dispersion of interstellar gas lasts for $5.5 \pm 1.1$\,Myr.

\section{PLANETARY-MASS OBJECTS}\label{sec:planemos}
One emerging sub-field in the study of nearby young associations is the discovery and characterization of substellar objects with extremely low masses, some of which are even estimated to have masses below the International Astronomical Union (IAU) working definition of a brown dwarf (above 13\,\mjup), below which even the fusion of deuterium is not believed to take place \citep{2022NewAR..9401641L}. Because these objects escape current definitions of both brown dwarfs and exoplanets (those require to be in orbit around a star by the same IAU working definition), they have generally been referred to as `isolated planetary-mass objects', or the contraction `planemos' for short. While these objects escape current definitions, it is likely that at least a fraction of them formed directly from the collapse of a molecular cloud fragment, similarly to how stars and brown dwarfs form (e.g., 2MASS~J1119$-$1137~AB; \citealp{2017ApJ...843L...4B}). Another plausible formation pathway for such objects is the formation within a multiple stellar system as a gas giant exoplanet, and the subsequent ejection due to instabilities driven by interactions with the host star system's orbit. Although such scenarios are believed to happen based on simulations \citep{2012MNRAS.422.1648V}, no direct evidence has yet been produced to conclusively show that one such isolated planetary-mass object formed in this manner.

While only a handful of planemos have been fully confirmed so far (e.g., see \citealp{2013ApJ...777L..20L,2015ApJ...808L..20G,2017ApJ...841L...1G,2023ApJ...943L..16S,2021ApJ...911....7Z}) including spectroscopic confirmations of their substellar nature and the measurement of their 3D velocities required to ensure the membership in their host young association is robust, a larger number of candidate planemos have been identified so far (e.g., see \citealp{2011ApJ...726...23G,2015ApJ...810..159M,2012ApJ...756...24S,2012ApJ...744..134M}), some of which now benefit from spectra thanks to the James Webb Space Telescope (JWST; see \citealp{2024AJ....167...19L}).

While planemos are notably hard to identify and confirm, the fact that they share many fundamental properties (masses, radii, temperatures, atmospheric structure and clouds) with directly-imaged giant exoplanets yet are not located in the immediate vicinity of a much brighter host star makes them valuable benchmarks to understand planetary atmospheres. These objects can be studied with spectroscopy at a resolution and signal-to-noise much higher than the typical gas giant exoplanets, meaning that their fundamental properties can be derived with a much greater accuracy.

Some recent examples include detailed studies of the planemos SIMP~J0136+0933 and 2MASS~J2139+0220 \citep{2023ApJ...944..138V}, which allowed determining that their observed photometric variability are caused by high-altitude patchy forsterite clouds, while deriving precise temperature-pressure profiles, atmospheric molecular abundances, cloud particle size distributions, and the identification of a deep iron cloud deck, similar to what has been observed for more massive brown dwarfs of similar effective temperatures, except enstatite was also found to be present in the atmospheres of the latest \citep{2017MNRAS.470.1177B,2021MNRAS.506.1944B}.

These are only previews of what will now be possible with the recent launch of the JWST, which is already slated for observations at a much wider range of wavelengths, and higher spectral resolutions and signal-to-noise, which will greatly help breaking the degeneracy between temperature-pressure profiles and non-gray atmospheric cloud opacity, while furthering the list of molecular gases which can be constrained at high precision. These future observations will serve as a guiding light to understand the atmospheres of exoplanets in the near future.


We thank the anonymous reviewer for their detailed comments that led to a substantial improvement of this work. We thank Leslie Moranta and Dominic Couture who provided help with generating some of the figures presented here.

\bibliographystyle{apj}


\begin{thebibliography}{}
\expandafter\ifx\csname natexlab\endcsname\relax\def\natexlab#1{#1}\fi

\bibitem[{{Adams}(1925)}]{1925PASP...37Q.158A}
{Adams}, W.~S. 1925, \pasp, 37, 158

\bibitem[{{Allende Prieto} {et~al.}(2013){Allende Prieto}, {Koesterke},
  {Ludwig}, {Freytag}, \& {Caffau}}]{2013AA...550A.103A}
{Allende Prieto}, C., {Koesterke}, L., {Ludwig}, H.~G., {Freytag}, B., \&
  {Caffau}, E. 2013, \aap, 550, A103

\bibitem[{{Anders} {et~al.}(2021){Anders}, {Cantat-Gaudin}, {Quadrino-Lodoso},
  {Gieles}, {Jordi}, {Castro-Ginard}, \&
  {Balaguer-N{\'u}{\~n}ez}}]{2021AA...645L...2A}
{Anders}, F., {Cantat-Gaudin}, T., {Quadrino-Lodoso}, I., {et~al.} 2021, \aap,
  645, L2

\bibitem[{{Artyukhina} \& {Kholopov}(1964)}]{1964SvA.....7..840A}
{Artyukhina}, N.~M., \& {Kholopov}, P.~N. 1964, \sovast, 7, 840

\bibitem[{{Asano} {et~al.}(2020){Asano}, {Fujii}, {Baba}, {B{\'e}dorf},
  {Sellentin}, \& {Portegies Zwart}}]{2020MNRAS.499.2416A}
{Asano}, T., {Fujii}, M.~S., {Baba}, J., {et~al.} 2020, \mnras, 499, 2416

\bibitem[{{Bailer-Jones} {et~al.}(2021){Bailer-Jones}, {Rybizki}, {Fouesneau},
  {Demleitner}, \& {Andrae}}]{2021AJ....161..147B}
{Bailer-Jones}, C.~A.~L., {Rybizki}, J., {Fouesneau}, M., {Demleitner}, M., \&
  {Andrae}, R. 2021, \aj, 161, 147

\bibitem[{{Bally}(2016)}]{2016ARAA..54..491B}
{Bally}, J. 2016, \araa, 54, 491

\bibitem[{{Barber} {et~al.}(2022){Barber}, {Mann}, {Bush}, {Tofflemire},
  {Kraus}, {Krolikowski}, {Vanderburg}, {Fields}, {Newton}, {Owens}, \&
  {Thao}}]{2022AJ....164...88B}
{Barber}, M.~G., {Mann}, A.~W., {Bush}, J.~L., {et~al.} 2022, \aj, 164, 88

\bibitem[{{Barnes}(2003)}]{2003ApJ...586..464B}
{Barnes}, S.~A. 2003, \apj, 586, 464

\bibitem[{{Barnes}(2009)}]{2009IAUS..258..345B}
{Barnes}, S.~A. 2009, in The Ages of Stars, ed. E.~E. {Mamajek}, D.~R.
  {Soderblom}, \& R.~F.~G. {Wyse}, Vol. 258, 345--356

\bibitem[{{Baroch} {et~al.}(2020){Baroch}, {Morales}, {Ribas}, {Herrero},
  {Rosich}, {Perger}, {Anglada-Escud{\'e}}, {Reiners}, {Caballero},
  {Quirrenbach}, {Amado}, {Jeffers}, {Cifuentes}, {Passegger}, {Schweitzer},
  {Lafarga}, {Bauer}, {B{\'e}jar}, {Colom{\'e}}, {Cort{\'e}s-Contreras},
  {Dreizler}, {Galad{\'\i}-Enr{\'\i}quez}, {Hatzes}, {Henning}, {Kaminski},
  {K{\"u}rster}, {Montes}, {Rodr{\'\i}guez-L{\'o}pez}, \&
  {Zechmeister}}]{2020AA...641A..69B}
{Baroch}, D., {Morales}, J.~C., {Ribas}, I., {et~al.} 2020, \aap, 641, A69

\bibitem[{{Basri} {et~al.}(1996){Basri}, {Marcy}, \&
  {Graham}}]{1996ApJ...458..600B}
{Basri}, G., {Marcy}, G.~W., \& {Graham}, J.~R. 1996, \apj, 458, 600

\bibitem[{{Best} {et~al.}(2017){Best}, {Liu}, {Dupuy}, \&
  {Magnier}}]{2017ApJ...843L...4B}
{Best}, W. M.~J., {Liu}, M.~C., {Dupuy}, T.~J., \& {Magnier}, E.~A. 2017,
  \apjl, 843, L4

\bibitem[{{Bhattacharya} {et~al.}(2022){Bhattacharya}, {Rao}, {Agarwal},
  {Balan}, \& {Vaidya}}]{2022MNRAS.517.3525B}
{Bhattacharya}, S., {Rao}, K.~K., {Agarwal}, M., {Balan}, S., \& {Vaidya}, K.
  2022, \mnras, 517, 3525

\bibitem[{{Blaauw}(1946)}]{1946PGro...52....1B}
{Blaauw}, A. 1946, Publications of the Kapteyn Astronomical Laboratory
  Groningen, 52, 1

\bibitem[{{Blaauw}(1964)}]{1964ARAA...2..213B}
---. 1964, \araa, 2, 213

\bibitem[{{Booth} {et~al.}(2017){Booth}, {Poppenhaeger}, {Watson}, {Silva
  Aguirre}, \& {Wolk}}]{2017MNRAS.471.1012B}
{Booth}, R.~S., {Poppenhaeger}, K., {Watson}, C.~A., {Silva Aguirre}, V., \&
  {Wolk}, S.~J. 2017, \mnras, 471, 1012

\bibitem[{{Brandt} \& {Huang}(2015)}]{2015ApJ...807...58B}
{Brandt}, T.~D., \& {Huang}, C.~X. 2015, \apj, 807, 58

\bibitem[{{Burke} {et~al.}(2004){Burke}, {Pinsonneault}, \&
  {Sills}}]{2004ApJ...604..272B}
{Burke}, C.~J., {Pinsonneault}, M.~H., \& {Sills}, A. 2004, \apj, 604, 272

\bibitem[{{Burningham} {et~al.}(2017){Burningham}, {Marley}, {Line}, {Lupu},
  {Visscher}, {Morley}, {Saumon}, \& {Freedman}}]{2017MNRAS.470.1177B}
{Burningham}, B., {Marley}, M.~S., {Line}, M.~R., {et~al.} 2017, \mnras, 470,
  1177

\bibitem[{{Burningham} {et~al.}(2021){Burningham}, {Faherty}, {Gonzales},
  {Marley}, {Visscher}, {Lupu}, {Gaarn}, {Fabienne Bieger}, {Freedman}, \&
  {Saumon}}]{2021MNRAS.506.1944B}
{Burningham}, B., {Faherty}, J.~K., {Gonzales}, E.~C., {et~al.} 2021, \mnras,
  506, 1944

\bibitem[{{Bus{\`a}} {et~al.}(2007){Bus{\`a}}, {Aznar Cuadrado}, {Terranegra},
  {Andretta}, \& {Gomez}}]{2007AA...466.1089B}
{Bus{\`a}}, I., {Aznar Cuadrado}, R., {Terranegra}, L., {Andretta}, V., \&
  {Gomez}, M.~T. 2007, \aap, 466, 1089

\bibitem[{Campello {et~al.}(2013)Campello, Moulavi, \&
  Sander}]{campello2013density}
Campello, R.~J., Moulavi, D., \& Sander, J. 2013, in Pacific-Asia conference on
  knowledge discovery and data mining, Springer, 160--172

\bibitem[{{Cantat-Gaudin} \& {Anders}(2020)}]{2020AA...633A..99C}
{Cantat-Gaudin}, T., \& {Anders}, F. 2020, \aap, 633, A99

\bibitem[{{Cargill} {et~al.}(2015){Cargill}, {Warren}, \&
  {Bradshaw}}]{2015RSPTA.37340260C}
{Cargill}, P.~J., {Warren}, H.~P., \& {Bradshaw}, S.~J. 2015, Philosophical
  Transactions of the Royal Society of London Series A, 373, 20140260

\bibitem[{{Castro-Ginard} {et~al.}(2020){Castro-Ginard}, {Jordi}, {Luri},
  {{\'A}lvarez Cid-Fuentes}, {Casamiquela}, {Anders}, {Cantat-Gaudin},
  {Mongui{\'o}}, {Balaguer-N{\'u}{\~n}ez}, {Sol{\`a}}, \&
  {Badia}}]{2020AA...635A..45C}
{Castro-Ginard}, A., {Jordi}, C., {Luri}, X., {et~al.} 2020, \aap, 635, A45

\bibitem[{{Chambers}(2021)}]{2021ApJ...914..102C}
{Chambers}, J. 2021, \apj, 914, 102

\bibitem[{{Chauvin} {et~al.}(2004){Chauvin}, {Lagrange}, {Dumas}, {Zuckerman},
  {Mouillet}, {Song}, {Beuzit}, \& {Lowrance}}]{2004AA...425L..29C}
{Chauvin}, G., {Lagrange}, A.~M., {Dumas}, C., {et~al.} 2004, \aap, 425, L29

\bibitem[{{Choi} {et~al.}(2016){Choi}, {Dotter}, {Conroy}, {Cantiello},
  {Paxton}, \& {Johnson}}]{2016ApJ...823..102C}
{Choi}, J., {Dotter}, A., {Conroy}, C., {et~al.} 2016, \apj, 823, 102

\bibitem[{{Christiansen} {et~al.}(2019){Christiansen}, {Beichman}, {Ciardi}, \&
  {Huber}}]{2019BAAS...51c.312C}
{Christiansen}, J., {Beichman}, C., {Ciardi}, D.~R., \& {Huber}, D. 2019,
  \baas, 51, 312

\bibitem[{{Couture} {et~al.}(2023){Couture}, {Gagn{\'e}}, \&
  {Doyon}}]{2023ApJ...946....6C}
{Couture}, D., {Gagn{\'e}}, J., \& {Doyon}, R. 2023, \apj, 946, 6

\bibitem[{{Crundall} {et~al.}(2019){Crundall}, {Ireland}, {Krumholz},
  {Federrath}, {{\v{Z}}erjal}, \& {Hansen}}]{2019MNRAS.489.3625C}
{Crundall}, T.~D., {Ireland}, M.~J., {Krumholz}, M.~R., {et~al.} 2019, \mnras,
  489, 3625

\bibitem[{{Currie} {et~al.}(2023){Currie}, {Biller}, {Lagrange}, {Marois},
  {Guyon}, {Nielsen}, {Bonnefoy}, \& {De Rosa}}]{2023ASPC..534..799C}
{Currie}, T., {Biller}, B., {Lagrange}, A., {et~al.} 2023, in Astronomical
  Society of the Pacific Conference Series, Vol. 534, Protostars and Planets
  VII, ed. S.~{Inutsuka}, Y.~{Aikawa}, T.~{Muto}, K.~{Tomida}, \& M.~{Tamura},
  799

\bibitem[{{Dai} {et~al.}(2019){Dai}, {Li}, \&
  {Stojkovic}}]{2019ApJ...871..119D}
{Dai}, D.-C., {Li}, Z., \& {Stojkovic}, D. 2019, \apj, 871, 119

\bibitem[{{David} {et~al.}(2021){David}, {Contardo}, {Sandoval}, {Angus}, {Lu},
  {Bedell}, {Curtis}, {Foreman-Mackey}, {Fulton}, {Grunblatt}, \&
  {Petigura}}]{2021AJ....161..265D}
{David}, T.~J., {Contardo}, G., {Sandoval}, A., {et~al.} 2021, \aj, 161, 265

\bibitem[{{de Bruijne}(1999)}]{1999MNRAS.306..381D}
{de Bruijne}, J. H.~J. 1999, \mnras, 306, 381

\bibitem[{{de Zeeuw} {et~al.}(1999){de Zeeuw}, {Hoogerwerf}, {de Bruijne},
  {Brown}, \& {Blaauw}}]{1999AJ....117..354D}
{de Zeeuw}, P.~T., {Hoogerwerf}, R., {de Bruijne}, J.~H.~J., {Brown}, A.~G.~A.,
  \& {Blaauw}, A. 1999, \aj, 117, 354

\bibitem[{{Dehnen} \& {Binney}(1998)}]{1998MNRAS.298..387D}
{Dehnen}, W., \& {Binney}, J.~J. 1998, \mnras, 298, 387

\bibitem[{{Dias} {et~al.}(2002){Dias}, {Alessi}, {Moitinho}, \&
  {L{\'e}pine}}]{2002AA...389..871D}
{Dias}, W.~S., {Alessi}, B.~S., {Moitinho}, A., \& {L{\'e}pine}, J.~R.~D. 2002,
  \aap, 389, 871

\bibitem[{{Dravins}(1987)}]{1987AA...172..200D}
{Dravins}, D. 1987, \aap, 172, 200

\bibitem[{{Dravins} {et~al.}(1981){Dravins}, {Lindegren}, \&
  {Nordlund}}]{1981AA....96..345D}
{Dravins}, D., {Lindegren}, L., \& {Nordlund}, A. 1981, \aap, 96, 345

\bibitem[{{Dravins} \& {Nordlund}(1990)}]{1990AA...228..203D}
{Dravins}, D., \& {Nordlund}, A. 1990, \aap, 228, 203

\bibitem[{{Duncan} {et~al.}(1991){Duncan}, {Vaughan}, {Wilson}, {Preston},
  {Frazer}, {Lanning}, {Misch}, {Mueller}, {Soyumer}, {Woodard}, {Baliunas},
  {Noyes}, {Hartmann}, {Porter}, {Zwaan}, {Middelkoop}, {Rutten}, \&
  {Mihalas}}]{1991ApJS...76..383D}
{Duncan}, D.~K., {Vaughan}, A.~H., {Wilson}, O.~C., {et~al.} 1991, \apjs, 76,
  383

\bibitem[{{Durrell} \& {Harris}(1993)}]{1993AJ....105.1420D}
{Durrell}, P.~R., \& {Harris}, W.~E. 1993, \aj, 105, 1420

\bibitem[{{Eberhard} \& {Schwarzschild}(1913)}]{1913ApJ....38..292E}
{Eberhard}, G., \& {Schwarzschild}, K. 1913, \apj, 38, 292

\bibitem[{{Eggen}(1958)}]{1958MNRAS.118..154E}
{Eggen}, O.~J. 1958, \mnras, 118, 154

\bibitem[{{Eldridge} \& {Stanway}(2022)}]{2022ARAA..60..455E}
{Eldridge}, J.~J., \& {Stanway}, E.~R. 2022, \araa, 60, 455

\bibitem[{{Elmegreen} \& {Clemens}(1985)}]{1985ApJ...294..523E}
{Elmegreen}, B.~G., \& {Clemens}, C. 1985, \apj, 294, 523

\bibitem[{{Engle}(2024)}]{2024ApJ...960...62E}
{Engle}, S.~G. 2024, \apj, 960, 62

\bibitem[{ESA(1997)}]{hipb}
ESA. 1997, in ESA Special Publication, Vol. 1200, ESA Special Publication

\bibitem[{{Ester} {et~al.}(1996){Ester}, {Kriegel}, {Sander}, \&
  {Xu}}]{1996kddm.conf..226E}
{Ester}, M., {Kriegel}, H.-P., {Sander}, J., \& {Xu}, X. 1996, in Second
  International Conference on Knowledge Discovery and Data Mining (KDD'96).
  Proceedings of a conference held August 2-4, 226--331

\bibitem[{{Fleming}(1998)}]{1998ApJ...504..461F}
{Fleming}, T.~A. 1998, \apj, 504, 461

\bibitem[{{Gagn{\'e}} {et~al.}(2015){Gagn{\'e}}, {Burgasser}, {Faherty},
  {Lafreni{\'e}re}, {Doyon}, {Filippazzo}, {Bowsher}, \&
  {Nicholls}}]{2015ApJ...808L..20G}
{Gagn{\'e}}, J., {Burgasser}, A.~J., {Faherty}, J.~K., {et~al.} 2015, \apjl,
  808, L20

\bibitem[{{Gagn{\'e}} \& {Faherty}(2018)}]{2018ApJ...862..138G}
{Gagn{\'e}}, J., \& {Faherty}, J.~K. 2018, \apj, 862, 138

\bibitem[{{Gagn{\'e}} {et~al.}(2021){Gagn{\'e}}, {Faherty}, {Moranta}, \&
  {Popinchalk}}]{2021ApJ...915L..29G}
{Gagn{\'e}}, J., {Faherty}, J.~K., {Moranta}, L., \& {Popinchalk}, M. 2021,
  \apjl, 915, L29

\bibitem[{{Gagn{\'e}} {et~al.}(2014){Gagn{\'e}}, {Lafreni{\`e}re}, {Doyon},
  {Malo}, \& {Artigau}}]{2014ApJ...783..121G}
{Gagn{\'e}}, J., {Lafreni{\`e}re}, D., {Doyon}, R., {Malo}, L., \& {Artigau},
  {\'E}. 2014, \apj, 783, 121

\bibitem[{{Gagn{\'e}} {et~al.}(2023){Gagn{\'e}}, {Moranta}, {Faherty}, {Kiman},
  {Couture}, {Larochelle}, {Popinchalk}, \& {Morrone}}]{2023ApJ...945..119G}
{Gagn{\'e}}, J., {Moranta}, L., {Faherty}, J.~K., {et~al.} 2023, \apj, 945, 119

\bibitem[{{Gagn{\'e}} {et~al.}(2018{\natexlab{a}}){Gagn{\'e}}, {Roy-Loubier},
  {Faherty}, {Doyon}, \& {Malo}}]{2018ApJ...860...43G}
{Gagn{\'e}}, J., {Roy-Loubier}, O., {Faherty}, J.~K., {Doyon}, R., \& {Malo},
  L. 2018{\natexlab{a}}, \apj, 860, 43

\bibitem[{{Gagn{\'e}} {et~al.}(2017){Gagn{\'e}}, {Faherty}, {Burgasser},
  {Artigau}, {Bouchard}, {Albert}, {Lafreni{\`e}re}, {Doyon}, \& {Bardalez
  Gagliuffi}}]{2017ApJ...841L...1G}
{Gagn{\'e}}, J., {Faherty}, J.~K., {Burgasser}, A.~J., {et~al.} 2017, \apjl,
  841, L1

\bibitem[{{Gagn{\'e}} {et~al.}(2018{\natexlab{b}}){Gagn{\'e}}, {Mamajek},
  {Malo}, {Riedel}, {Rodriguez}, {Lafreni{\`e}re}, {Faherty}, {Roy-Loubier},
  {Pueyo}, {Robin}, \& {Doyon}}]{2018ApJ...856...23G}
{Gagn{\'e}}, J., {Mamajek}, E.~E., {Malo}, L., {et~al.} 2018{\natexlab{b}},
  \apj, 856, 23

\bibitem[{{Gaia Collaboration} {et~al.}(2016){Gaia Collaboration}, {Prusti},
  {de Bruijne}, {Brown}, {Vallenari}, {Babusiaux}, {Bailer-Jones}, {Bastian},
  {Biermann}, {Evans}, {Eyer}, {Jansen}, {Jordi}, {Klioner}, {Lammers},
  {Lindegren}, {Luri}, {Mignard}, {Milligan}, {Panem}, {Poinsignon},
  {Pourbaix}, {Randich}, {Sarri}, {Sartoretti}, {Siddiqui}, {Soubiran},
  {Valette}, {van Leeuwen}, {Walton}, {Aerts}, {Arenou}, {Cropper}, {Drimmel},
  {H{\o}g}, {Katz}, {Lattanzi}, {O'Mullane}, {Grebel}, {Holland}, {Huc},
  {Passot}, {Bramante}, {Cacciari}, {Casta{\~n}eda}, {Chaoul}, {Cheek}, {De
  Angeli}, {Fabricius}, {Guerra}, {Hern{\'a}ndez}, {Jean-Antoine-Piccolo},
  {Masana}, {Messineo}, {Mowlavi}, {Nienartowicz}, {Ord{\'o}{\~n}ez-Blanco},
  {Panuzzo}, {Portell}, {Richards}, {Riello}, {Seabroke}, {Tanga},
  {Th{\'e}venin}, {Torra}, {Els}, {Gracia-Abril}, {Comoretto},
  {Garcia-Reinaldos}, {Lock}, {Mercier}, {Altmann}, {Andrae}, {Astraatmadja},
  {Bellas-Velidis}, {Benson}, {Berthier}, {Blomme}, {Busso}, {Carry},
  {Cellino}, {Clementini}, {Cowell}, {Creevey}, {Cuypers}, {Davidson}, {De
  Ridder}, {de Torres}, {Delchambre}, {Dell'Oro}, {Ducourant}, {Fr{\'e}mat},
  {Garc{\'\i}a-Torres}, {Gosset}, {Halbwachs}, {Hambly}, {Harrison}, {Hauser},
  {Hestroffer}, {Hodgkin}, {Huckle}, {Hutton}, {Jasniewicz}, {Jordan},
  {Kontizas}, {Korn}, {Lanzafame}, {Manteiga}, {Moitinho}, {Muinonen},
  {Osinde}, {Pancino}, {Pauwels}, {Petit}, {Recio-Blanco}, {Robin}, {Sarro},
  {Siopis}, {Smith}, {Smith}, {Sozzetti}, {Thuillot}, {van Reeven}, {Viala},
  {Abbas}, {Abreu Aramburu}, {Accart}, {Aguado}, {Allan}, {Allasia},
  {Altavilla}, {{\'A}lvarez}, {Alves}, {Anderson}, {Andrei}, {Anglada Varela},
  {Antiche}, {Antoja}, {Ant{\'o}n}, {Arcay}, {Atzei}, {Ayache}, {Bach},
  {Baker}, {Balaguer-N{\'u}{\~n}ez}, {Barache}, {Barata}, {Barbier}, {Barblan},
  {Baroni}, {Barrado y Navascu{\'e}s}, {Barros}, {Barstow}, {Becciani},
  {Bellazzini}, {Bellei}, {Bello Garc{\'\i}a}, {Belokurov}, {Bendjoya},
  {Berihuete}, {Bianchi}, {Bienaym{\'e}}, {Billebaud}, {Blagorodnova},
  {Blanco-Cuaresma}, {Boch}, {Bombrun}, {Borrachero}, {Bouquillon}, {Bourda},
  {Bouy}, {Bragaglia}, {Breddels}, {Brouillet}, {Br{\"u}semeister},
  {Bucciarelli}, {Budnik}, {Burgess}, {Burgon}, {Burlacu}, {Busonero}, {Buzzi},
  {Caffau}, {Cambras}, {Campbell}, {Cancelliere}, {Cantat-Gaudin}, {Carlucci},
  {Carrasco}, {Castellani}, {Charlot}, {Charnas}, {Charvet}, {Chassat},
  {Chiavassa}, {Clotet}, {Cocozza}, {Collins}, {Collins}, {Costigan}, {Crifo},
  {Cross}, {Crosta}, {Crowley}, {Dafonte}, {Damerdji}, {Dapergolas}, {David},
  {David}, {De Cat}, {de Felice}, {de Laverny}, {De Luise}, {De March}, {de
  Martino}, {de Souza}, {Debosscher}, {del Pozo}, {Delbo}, {Delgado},
  {Delgado}, {di Marco}, {Di Matteo}, {Diakite}, {Distefano}, {Dolding}, {Dos
  Anjos}, {Drazinos}, {Dur{\'a}n}, {Dzigan}, {Ecale}, {Edvardsson}, {Enke},
  {Erdmann}, {Escolar}, {Espina}, {Evans}, {Eynard Bontemps}, {Fabre},
  {Fabrizio}, {Faigler}, {Falc{\~a}o}, {Farr{\`a}s Casas}, {Faye}, {Federici},
  {Fedorets}, {Fern{\'a}ndez-Hern{\'a}ndez}, {Fernique}, {Fienga}, {Figueras},
  {Filippi}, {Findeisen}, {Fonti}, {Fouesneau}, {Fraile}, {Fraser}, {Fuchs},
  {Furnell}, {Gai}, {Galleti}, {Galluccio}, {Garabato}, {Garc{\'\i}a-Sedano},
  {Gar{\'e}}, {Garofalo}, {Garralda}, {Gavras}, {Gerssen}, {Geyer}, {Gilmore},
  {Girona}, {Giuffrida}, {Gomes}, {Gonz{\'a}lez-Marcos},
  {Gonz{\'a}lez-N{\'u}{\~n}ez}, {Gonz{\'a}lez-Vidal}, {Granvik}, {Guerrier},
  {Guillout}, {Guiraud}, {G{\'u}rpide}, {Guti{\'e}rrez-S{\'a}nchez}, {Guy},
  {Haigron}, {Hatzidimitriou}, {Haywood}, {Heiter}, {Helmi}, {Hobbs},
  {Hofmann}, {Holl}, {Holland}, {Hunt}, {Hypki}, {Icardi}, {Irwin}, {Jevardat
  de Fombelle}, {Jofr{\'e}}, {Jonker}, {Jorissen}, {Julbe}, {Karampelas},
  {Kochoska}, {Kohley}, {Kolenberg}, {Kontizas}, {Koposov}, {Kordopatis},
  {Koubsky}, {Kowalczyk}, {Krone-Martins}, {Kudryashova}, {Kull}, {Bachchan},
  {Lacoste-Seris}, {Lanza}, {Lavigne}, {Le Poncin-Lafitte}, {Lebreton},
  {Lebzelter}, {Leccia}, {Leclerc}, {Lecoeur-Taibi}, {Lemaitre}, {Lenhardt},
  {Leroux}, {Liao}, {Licata}, {Lindstr{\o}m}, {Lister}, {Livanou}, {Lobel},
  {L{\"o}ffler}, {L{\'o}pez}, {Lopez-Lozano}, {Lorenz}, {Loureiro},
  {MacDonald}, {Magalh{\~a}es Fernandes}, {Managau}, {Mann}, {Mantelet},
  {Marchal}, {Marchant}, {Marconi}, {Marie}, {Marinoni}, {Marrese},
  {Marschalk{\'o}}, {Marshall}, {Mart{\'\i}n-Fleitas}, {Martino}, {Mary},
  {Matijevi{\v{c}}}, {Mazeh}, {McMillan}, {Messina}, {Mestre}, {Michalik},
  {Millar}, {Miranda}, {Molina}, {Molinaro}, {Molinaro}, {Moln{\'a}r},
  {Moniez}, {Montegriffo}, {Monteiro}, {Mor}, {Mora}, {Morbidelli}, {Morel},
  {Morgenthaler}, {Morley}, {Morris}, {Mulone}, {Muraveva}, {Musella},
  {Narbonne}, {Nelemans}, {Nicastro}, {Noval}, {Ord{\'e}novic},
  {Ordieres-Mer{\'e}}, {Osborne}, {Pagani}, {Pagano}, {Pailler}, {Palacin},
  {Palaversa}, {Parsons}, {Paulsen}, {Pecoraro}, {Pedrosa}, {Pentik{\"a}inen},
  {Pereira}, {Pichon}, {Piersimoni}, {Pineau}, {Plachy}, {Plum}, {Poujoulet},
  {Pr{\v{s}}a}, {Pulone}, {Ragaini}, {Rago}, {Rambaux}, {Ramos-Lerate},
  {Ranalli}, {Rauw}, {Read}, {Regibo}, {Renk}, {Reyl{\'e}}, {Ribeiro},
  {Rimoldini}, {Ripepi}, {Riva}, {Rixon}, {Roelens}, {Romero-G{\'o}mez},
  {Rowell}, {Royer}, {Rudolph}, {Ruiz-Dern}, {Sadowski}, {Sagrist{\`a}
  Sell{\'e}s}, {Sahlmann}, {Salgado}, {Salguero}, {Sarasso}, {Savietto},
  {Schnorhk}, {Schultheis}, {Sciacca}, {Segol}, {Segovia}, {Segransan},
  {Serpell}, {Shih}, {Smareglia}, {Smart}, {Smith}, {Solano}, {Solitro},
  {Sordo}, {Soria Nieto}, {Souchay}, {Spagna}, {Spoto}, {Stampa}, {Steele},
  {Steidelm{\"u}ller}, {Stephenson}, {Stoev}, {Suess}, {S{\"u}veges}, {Surdej},
  {Szabados}, {Szegedi-Elek}, {Tapiador}, {Taris}, {Tauran}, {Taylor},
  {Teixeira}, {Terrett}, {Tingley}, {Trager}, {Turon}, {Ulla}, {Utrilla},
  {Valentini}, {van Elteren}, {Van Hemelryck}, {van Leeuwen}, {Varadi},
  {Vecchiato}, {Veljanoski}, {Via}, {Vicente}, {Vogt}, {Voss}, {Votruba},
  {Voutsinas}, {Walmsley}, {Weiler}, {Weingrill}, {Werner}, {Wevers},
  {Whitehead}, {Wyrzykowski}, {Yoldas}, {{\v{Z}}erjal}, {Zucker}, {Zurbach},
  {Zwitter}, {Alecu}, {Allen}, {Allende Prieto}, {Amorim},
  {Anglada-Escud{\'e}}, {Arsenijevic}, {Azaz}, {Balm}, {Beck}, {Bernstein},
  {Bigot}, {Bijaoui}, {Blasco}, {Bonfigli}, {Bono}, {Boudreault}, {Bressan},
  {Brown}, {Brunet}, {Bunclark}, {Buonanno}, {Butkevich}, {Carret}, {Carrion},
  {Chemin}, {Ch{\'e}reau}, {Corcione}, {Darmigny}, {de Boer}, {de Teodoro}, {de
  Zeeuw}, {Delle Luche}, {Domingues}, {Dubath}, {Fodor}, {Fr{\'e}zouls},
  {Fries}, {Fustes}, {Fyfe}, {Gallardo}, {Gallegos}, {Gardiol}, {Gebran},
  {Gomboc}, {G{\'o}mez}, {Grux}, {Gueguen}, {Heyrovsky}, {Hoar}, {Iannicola},
  {Isasi Parache}, {Janotto}, {Joliet}, {Jonckheere}, {Keil}, {Kim},
  {Klagyivik}, {Klar}, {Knude}, {Kochukhov}, {Kolka}, {Kos}, {Kutka}, {Lainey},
  {LeBouquin}, {Liu}, {Loreggia}, {Makarov}, {Marseille}, {Martayan},
  {Martinez-Rubi}, {Massart}, {Meynadier}, {Mignot}, {Munari}, {Nguyen},
  {Nordlander}, {Ocvirk}, {O'Flaherty}, {Olias Sanz}, {Ortiz}, {Osorio},
  {Oszkiewicz}, {Ouzounis}, {Palmer}, {Park}, {Pasquato}, {Peltzer}, {Peralta},
  {P{\'e}turaud}, {Pieniluoma}, {Pigozzi}, {Poels}, {Prat}, {Prod'homme},
  {Raison}, {Rebordao}, {Risquez}, {Rocca-Volmerange}, {Rosen}, {Ruiz-Fuertes},
  {Russo}, {Sembay}, {Serraller Vizcaino}, {Short}, {Siebert}, {Silva},
  {Sinachopoulos}, {Slezak}, {Soffel}, {Sosnowska}, {Strai{\v{z}}ys}, {ter
  Linden}, {Terrell}, {Theil}, {Tiede}, {Troisi}, {Tsalmantza}, {Tur},
  {Vaccari}, {Vachier}, {Valles}, {Van Hamme}, {Veltz}, {Virtanen}, {Wallut},
  {Wichmann}, {Wilkinson}, {Ziaeepour}, \& {Zschocke}}]{2016AA...595A...1G}
{Gaia Collaboration}, {Prusti}, T., {de Bruijne}, J.~H.~J., {et~al.} 2016,
  \aap, 595, A1

\bibitem[{{Gaia Collaboration} {et~al.}(2021){Gaia Collaboration}, {Smart},
  {Sarro}, {Rybizki}, {Reyl{\'e}}, {Robin}, {Hambly}, {Abbas}, {Barstow}, {de
  Bruijne}, {Bucciarelli}, {Carrasco}, {Cooper}, {Hodgkin}, {Masana},
  {Michalik}, {Sahlmann}, {Sozzetti}, {Brown}, {Vallenari}, {Prusti},
  {Babusiaux}, {Biermann}, {Creevey}, {Evans}, {Eyer}, {Hutton}, {Jansen},
  {Jordi}, {Klioner}, {Lammers}, {Lindegren}, {Luri}, {Mignard}, {Panem},
  {Pourbaix}, {Randich}, {Sartoretti}, {Soubiran}, {Walton}, {Arenou},
  {Bailer-Jones}, {Bastian}, {Cropper}, {Drimmel}, {Katz}, {Lattanzi}, {van
  Leeuwen}, {Bakker}, {Casta{\~n}eda}, {De Angeli}, {Ducourant}, {Fabricius},
  {Fouesneau}, {Fr{\'e}mat}, {Guerra}, {Guerrier}, {Guiraud}, {Jean-Antoine
  Piccolo}, {Messineo}, {Mowlavi}, {Nicolas}, {Nienartowicz}, {Pailler},
  {Panuzzo}, {Riclet}, {Roux}, {Seabroke}, {Sordo}, {Tanga}, {Th{\'e}venin},
  {Gracia-Abril}, {Portell}, {Teyssier}, {Altmann}, {Andrae}, {Bellas-Velidis},
  {Benson}, {Berthier}, {Blomme}, {Brugaletta}, {Burgess}, {Busso}, {Carry},
  {Cellino}, {Cheek}, {Clementini}, {Damerdji}, {Davidson}, {Delchambre},
  {Dell'Oro}, {Fern{\'a}ndez-Hern{\'a}ndez}, {Galluccio}, {Garc{\'\i}a-Lario},
  {Garcia-Reinaldos}, {Gonz{\'a}lez-N{\'u}{\~n}ez}, {Gosset}, {Haigron},
  {Halbwachs}, {Harrison}, {Hatzidimitriou}, {Heiter}, {Hern{\'a}ndez},
  {Hestroffer}, {Holl}, {Jan{\ss}en}, {Jevardat de Fombelle}, {Jordan},
  {Krone-Martins}, {Lanzafame}, {L{\"o}ffler}, {Lorca}, {Manteiga}, {Marchal},
  {Marrese}, {Moitinho}, {Mora}, {Muinonen}, {Osborne}, {Pancino}, {Pauwels},
  {Recio-Blanco}, {Richards}, {Riello}, {Rimoldini}, {Roegiers}, {Siopis},
  {Smith}, {Ulla}, {Utrilla}, {van Leeuwen}, {van Reeven}, {Abreu Aramburu},
  {Accart}, {Aerts}, {Aguado}, {Ajaj}, {Altavilla}, {{\'A}lvarez}, {{\'A}lvarez
  Cid-Fuentes}, {Alves}, {Anderson}, {Anglada Varela}, {Antoja}, {Audard},
  {Baines}, {Baker}, {Balaguer-N{\'u}{\~n}ez}, {Balbinot}, {Balog}, {Barache},
  {Barbato}, {Barros}, {Bartolom{\'e}}, {Bassilana}, {Bauchet},
  {Baudesson-Stella}, {Becciani}, {Bellazzini}, {Bernet}, {Bertone}, {Bianchi},
  {Blanco-Cuaresma}, {Boch}, {Bombrun}, {Bossini}, {Bouquillon}, {Bragaglia},
  {Bramante}, {Breedt}, {Bressan}, {Brouillet}, {Burlacu}, {Busonero},
  {Butkevich}, {Buzzi}, {Caffau}, {Cancelliere}, {C{\'a}novas},
  {Cantat-Gaudin}, {Carballo}, {Carlucci}, {Carnerero}, {Casamiquela},
  {Castellani}, {Castro-Ginard}, {Castro Sampol}, {Chaoul}, {Charlot},
  {Chemin}, {Chiavassa}, {Cioni}, {Comoretto}, {Cornez}, {Cowell}, {Crifo},
  {Crosta}, {Crowley}, {Dafonte}, {Dapergolas}, {David}, {David}, {de Laverny},
  {De Luise}, {De March}, {De Ridder}, {de Souza}, {de Teodoro}, {de Torres},
  {del Peloso}, {del Pozo}, {Delgado}, {Delgado}, {Delisle}, {Di Matteo},
  {Diakite}, {Diener}, {Distefano}, {Dolding}, {Eappachen}, {Edvardsson},
  {Enke}, {Esquej}, {Fabre}, {Fabrizio}, {Faigler}, {Fedorets}, {Fernique},
  {Fienga}, {Figueras}, {Fouron}, {Fragkoudi}, {Fraile}, {Franke}, {Gai},
  {Garabato}, {Garcia-Gutierrez}, {Garc{\'\i}a-Torres}, {Garofalo}, {Gavras},
  {Gerlach}, {Geyer}, {Giacobbe}, {Gilmore}, {Girona}, {Giuffrida}, {Gomel},
  {Gomez}, {Gonzalez-Santamaria}, {Gonz{\'a}lez-Vidal}, {Granvik},
  {Guti{\'e}rrez-S{\'a}nchez}, {Guy}, {Hauser}, {Haywood}, {Helmi}, {Hidalgo},
  {Hilger}, {H{\l}adczuk}, {Hobbs}, {Holland}, {Huckle}, {Jasniewicz},
  {Jonker}, {Juaristi Campillo}, {Julbe}, {Karbevska}, {Kervella}, {Khanna},
  {Kochoska}, {Kontizas}, {Kordopatis}, {Korn}, {Kostrzewa-Rutkowska},
  {Kruszy{\'n}ska}, {Lambert}, {Lanza}, {Lasne}, {Le Campion}, {Le Fustec},
  {Lebreton}, {Lebzelter}, {Leccia}, {Leclerc}, {Lecoeur-Taibi}, {Liao},
  {Licata}, {Lindstr{\o}m}, {Lister}, {Livanou}, {Lobel}, {Madrero Pardo},
  {Managau}, {Mann}, {Marchant}, {Marconi}, {Marcos Santos}, {Marinoni},
  {Marocco}, {Marshall}, {Martin Polo}, {Mart{\'\i}n-Fleitas}, {Masip},
  {Massari}, {Mastrobuono-Battisti}, {Mazeh}, {McMillan}, {Messina}, {Millar},
  {Mints}, {Molina}, {Molinaro}, {Moln{\'a}r}, {Montegriffo}, {Mor},
  {Morbidelli}, {Morel}, {Morris}, {Mulone}, {Munoz}, {Muraveva}, {Murphy},
  {Musella}, {Noval}, {Ord{\'e}novic}, {Orr{\`u}}, {Osinde}, {Pagani},
  {Pagano}, {Palaversa}, {Palicio}, {Panahi}, {Pawlak}, {Pe{\~n}alosa
  Esteller}, {Penttil{\"a}}, {Piersimoni}, {Pineau}, {Plachy}, {Plum},
  {Poggio}, {Poretti}, {Poujoulet}, {Pr{\v{s}}a}, {Pulone}, {Racero},
  {Ragaini}, {Rainer}, {Raiteri}, {Rambaux}, {Ramos}, {Ramos-Lerate}, {Re
  Fiorentin}, {Regibo}, {Ripepi}, {Riva}, {Rixon}, {Robichon}, {Robin},
  {Roelens}, {Rohrbasser}, {Romero-G{\'o}mez}, {Rowell}, {Royer}, {Rybicki},
  {Sadowski}, {Sagrist{\`a} Sell{\'e}s}, {Salgado}, {Salguero}, {Samaras},
  {Sanchez Gimenez}, {Sanna}, {Santove{\~n}a}, {Sarasso}, {Schultheis},
  {Sciacca}, {Segol}, {Segovia}, {S{\'e}gransan}, {Semeux}, {Shahaf},
  {Siddiqui}, {Siebert}, {Siltala}, {Slezak}, {Solano}, {Solitro}, {Souami},
  {Souchay}, {Spagna}, {Spoto}, {Steele}, {Steidelm{\"u}ller}, {Stephenson},
  {S{\"u}veges}, {Szabados}, {Szegedi-Elek}, {Taris}, {Tauran}, {Taylor},
  {Teixeira}, {Thuillot}, {Tonello}, {Torra}, {Torra}, {Turon}, {Unger},
  {Vaillant}, {van Dillen}, {Vanel}, {Vecchiato}, {Viala}, {Vicente},
  {Voutsinas}, {Weiler}, {Wevers}, {Wyrzykowski}, {Yoldas}, {Yvard}, {Zhao},
  {Zorec}, {Zucker}, {Zurbach}, \& {Zwitter}}]{2021AA...649A...6G}
{Gaia Collaboration}, {Smart}, R.~L., {Sarro}, L.~M., {et~al.} 2021, \aap, 649,
  A6

\bibitem[{{Galindo-Guil} {et~al.}(2022){Galindo-Guil}, {Barrado}, {Bouy},
  {Olivares}, {Bayo}, {Morales-Calder{\'o}n}, {Hu{\'e}lamo}, {Sarro},
  {Rivi{\`e}re-Marichalar}, {Stoev}, {Montesinos}, \&
  {Stauffer}}]{2022AA...664A..70G}
{Galindo-Guil}, F.~J., {Barrado}, D., {Bouy}, H., {et~al.} 2022, \aap, 664, A70

\bibitem[{{Gallet} \& {Bouvier}(2013)}]{2013AA...556A..36G}
{Gallet}, F., \& {Bouvier}, J. 2013, \aap, 556, A36

\bibitem[{{Geers} {et~al.}(2011){Geers}, {Scholz}, {Jayawardhana}, {Lee},
  {Lafreni{\`e}re}, \& {Tamura}}]{2011ApJ...726...23G}
{Geers}, V., {Scholz}, A., {Jayawardhana}, R., {et~al.} 2011, \apj, 726, 23

\bibitem[{{Gold}(1965)}]{1965IAUS...22..390G}
{Gold}, T. 1965, in Stellar and Solar Magnetic Fields, ed. R.~{Lust}, Vol.~22,
  390

\bibitem[{{Gray}(1982)}]{1982ApJ...255..200G}
{Gray}, D.~F. 1982, \apj, 255, 200

\bibitem[{{Gray}(2010)}]{2010ApJ...721..670G}
---. 2010, \apj, 721, 670

\bibitem[{{Griffin}(1982)}]{1982JApA....3..383G}
{Griffin}, R.~F. 1982, Journal of Astrophysics and Astronomy, 3, 383

\bibitem[{{Gunn} {et~al.}(1988){Gunn}, {Griffin}, {Griffin}, \&
  {Zimmerman}}]{1988AJ.....96..198G}
{Gunn}, J.~E., {Griffin}, R.~F., {Griffin}, R.~E.~M., \& {Zimmerman}, B.~A.
  1988, \aj, 96, 198

\bibitem[{{Hale} \& {Ellerman}(1920)}]{1920PASP...32..272H}
{Hale}, G.~E., \& {Ellerman}, F. 1920, \pasp, 32, 272

\bibitem[{{Hamilton} \& {Lester}(1999)}]{1999PASP..111.1132H}
{Hamilton}, D., \& {Lester}, J.~B. 1999, \pasp, 111, 1132

\bibitem[{{Hatzes}(2016)}]{2016ASSL..428....3H}
{Hatzes}, A.~P. 2016, in Astrophysics and Space Science Library, Vol. 428,
  Methods of Detecting Exoplanets: 1st Advanced School on Exoplanetary Science,
  ed. V.~{Bozza}, L.~{Mancini}, \& A.~{Sozzetti}, 3

\bibitem[{{Hempelmann} {et~al.}(1996){Hempelmann}, {Schmitt}, \&
  {St{\c{e}}pie{\'n}}}]{1996AA...305..284H}
{Hempelmann}, A., {Schmitt}, J.~H.~M.~M., \& {St{\c{e}}pie{\'n}}, K. 1996,
  \aap, 305, 284

\bibitem[{{Hentschel}(1994)}]{1994AHES...47..143H}
{Hentschel}, K. 1994, Archive for History of Exact Sciences, 47, 143

\bibitem[{{Jerabkova} {et~al.}(2019){Jerabkova}, {Boffin}, {Beccari}, \&
  {Anderson}}]{2019MNRAS.489.4418J}
{Jerabkova}, T., {Boffin}, H. M.~J., {Beccari}, G., \& {Anderson}, R.~I. 2019,
  \mnras, 489, 4418

\bibitem[{{Jerabkova} {et~al.}(2021){Jerabkova}, {Boffin}, {Beccari}, {de
  Marchi}, {de Bruijne}, \& {Prusti}}]{2021AA...647A.137J}
{Jerabkova}, T., {Boffin}, H. M.~J., {Beccari}, G., {et~al.} 2021, \aap, 647,
  A137

\bibitem[{{Johnson} \& {Soderblom}(1987)}]{1987AJ.....93..864J}
{Johnson}, D. R.~H., \& {Soderblom}, D.~R. 1987, \aj, 93, 864

\bibitem[{{Kastner} {et~al.}(1997){Kastner}, {Zuckerman}, {Weintraub}, \&
  {Forveille}}]{1997Sci...277...67K}
{Kastner}, J.~H., {Zuckerman}, B., {Weintraub}, D.~A., \& {Forveille}, T. 1997,
  Science, 277, 67

\bibitem[{{Kawaler}(1988)}]{1988ApJ...333..236K}
{Kawaler}, S.~D. 1988, \apj, 333, 236

\bibitem[{{Kerr} {et~al.}(2022{\natexlab{a}}){Kerr}, {Kraus}, {Murphy},
  {Krolikowski}, {Bedding}, \& {Rizzuto}}]{2022ApJ...941..143K}
{Kerr}, R., {Kraus}, A.~L., {Murphy}, S.~J., {et~al.} 2022{\natexlab{a}}, \apj,
  941, 143

\bibitem[{{Kerr} {et~al.}(2022{\natexlab{b}}){Kerr}, {Kraus}, {Murphy},
  {Krolikowski}, {Offner}, {Tofflemire}, \& {Rizzuto}}]{2022ApJ...941...49K}
---. 2022{\natexlab{b}}, \apj, 941, 49

\bibitem[{{Kerr} {et~al.}(2023){Kerr}, {Kraus}, \&
  {Rizzuto}}]{2023ApJ...954..134K}
{Kerr}, R., {Kraus}, A.~L., \& {Rizzuto}, A.~C. 2023, \apj, 954, 134

\bibitem[{{Kerr} {et~al.}(2021){Kerr}, {Rizzuto}, {Kraus}, \&
  {Offner}}]{2021ApJ...917...23K}
{Kerr}, R. M.~P., {Rizzuto}, A.~C., {Kraus}, A.~L., \& {Offner}, S. S.~R. 2021,
  \apj, 917, 23

\bibitem[{{Kim} {et~al.}(2021){Kim}, {Chevance}, {Kruijssen}, {Schruba},
  {Sandstrom}, {Barnes}, {Bigiel}, {Blanc}, {Cao}, {Dale}, {Faesi}, {Glover},
  {Grasha}, {Groves}, {Herrera}, {Klessen}, {Kreckel}, {Lee}, {Leroy}, {Pety},
  {Querejeta}, {Schinnerer}, {Sun}, {Usero}, {Ward}, \&
  {Williams}}]{2021MNRAS.504..487K}
{Kim}, J., {Chevance}, M., {Kruijssen}, J.~M.~D., {et~al.} 2021, \mnras, 504,
  487

\bibitem[{{Kounkel} \& {Covey}(2019)}]{2019AJ....158..122K}
{Kounkel}, M., \& {Covey}, K. 2019, \aj, 158, 122

\bibitem[{{Lada} \& {Lada}(2003)}]{2003ARAA..41...57L}
{Lada}, C.~J., \& {Lada}, E.~A. 2003, \araa, 41, 57

\bibitem[{{Lada} {et~al.}(1984){Lada}, {Margulis}, \&
  {Dearborn}}]{1984ApJ...285..141L}
{Lada}, C.~J., {Margulis}, M., \& {Dearborn}, D. 1984, \apj, 285, 141

\bibitem[{{Lamers} {et~al.}(2005){Lamers}, {Gieles}, {Bastian}, {Baumgardt},
  {Kharchenko}, \& {Portegies Zwart}}]{2005AA...441..117L}
{Lamers}, H.~J.~G.~L.~M., {Gieles}, M., {Bastian}, N., {et~al.} 2005, \aap,
  441, 117

\bibitem[{{Laughlin} {et~al.}(1997){Laughlin}, {Bodenheimer}, \&
  {Adams}}]{1997ApJ...482..420L}
{Laughlin}, G., {Bodenheimer}, P., \& {Adams}, F.~C. 1997, \apj, 482, 420

\bibitem[{{Le{\~a}o} {et~al.}(2019){Le{\~a}o}, {Pasquini}, {Ludwig}, \& {de
  Medeiros}}]{2019MNRAS.483.5026L}
{Le{\~a}o}, I.~C., {Pasquini}, L., {Ludwig}, H.~G., \& {de Medeiros}, J.~R.
  2019, \mnras, 483, 5026

\bibitem[{{Lecavelier des Etangs} \& {Lissauer}(2022)}]{2022NewAR..9401641L}
{Lecavelier des Etangs}, A., \& {Lissauer}, J.~J. 2022, \nar, 94, 101641

\bibitem[{{L{\'e}ger} {et~al.}(2009){L{\'e}ger}, {Rouan}, {Schneider}, {Barge},
  {Fridlund}, {Samuel}, {Ollivier}, {Guenther}, {Deleuil}, {Deeg}, {Auvergne},
  {Alonso}, {Aigrain}, {Alapini}, {Almenara}, {Baglin}, {Barbieri}, {Bruntt},
  {Bord{\'e}}, {Bouchy}, {Cabrera}, {Catala}, {Carone}, {Carpano}, {Csizmadia},
  {Dvorak}, {Erikson}, {Ferraz-Mello}, {Foing}, {Fressin}, {Gandolfi},
  {Gillon}, {Gondoin}, {Grasset}, {Guillot}, {Hatzes}, {H{\'e}brard}, {Jorda},
  {Lammer}, {Llebaria}, {Loeillet}, {Mayor}, {Mazeh}, {Moutou}, {P{\"a}tzold},
  {Pont}, {Queloz}, {Rauer}, {Renner}, {Samadi}, {Shporer}, {Sotin}, {Tingley},
  {Wuchterl}, {Adda}, {Agogu}, {Appourchaux}, {Ballans}, {Baron}, {Beaufort},
  {Bellenger}, {Berlin}, {Bernardi}, {Blouin}, {Baudin}, {Bodin}, {Boisnard},
  {Boit}, {Bonneau}, {Borzeix}, {Briet}, {Buey}, {Butler}, {Cailleau},
  {Cautain}, {Chabaud}, {Chaintreuil}, {Chiavassa}, {Costes}, {Cuna Parrho},
  {de Oliveira Fialho}, {Decaudin}, {Defise}, {Djalal}, {Epstein}, {Exil},
  {Faur{\'e}}, {Fenouillet}, {Gaboriaud}, {Gallic}, {Gamet}, {Gavalda},
  {Grolleau}, {Gruneisen}, {Gueguen}, {Guis}, {Guivarc'h}, {Guterman},
  {Hallouard}, {Hasiba}, {Heuripeau}, {Huntzinger}, {Hustaix}, {Imad},
  {Imbert}, {Johlander}, {Jouret}, {Journoud}, {Karioty}, {Kerjean},
  {Lafaille}, {Lafond}, {Lam-Trong}, {Landiech}, {Lapeyrere}, {Larqu{\'e}},
  {Laudet}, {Lautier}, {Lecann}, {Lefevre}, {Leruyet}, {Levacher}, {Magnan},
  {Mazy}, {Mertens}, {Mesnager}, {Meunier}, {Michel}, {Monjoin}, {Naudet},
  {Nguyen-Kim}, {Orcesi}, {Ottacher}, {Perez}, {Peter}, {Plasson}, {Plesseria},
  {Pontet}, {Pradines}, {Quentin}, {Reynaud}, {Rolland}, {Rollenhagen},
  {Romagnan}, {Russ}, {Schmidt}, {Schwartz}, {Sebbag}, {Sedes}, {Smit},
  {Steller}, {Sunter}, {Surace}, {Tello}, {Tiph{\`e}ne}, {Toulouse}, {Ulmer},
  {Vandermarcq}, {Vergnault}, {Vuillemin}, \& {Zanatta}}]{2009AA...506..287L}
{L{\'e}ger}, A., {Rouan}, D., {Schneider}, J., {et~al.} 2009, \aap, 506, 287

\bibitem[{{Levine}(1974)}]{1974ApJ...190..447L}
{Levine}, R.~H. 1974, \apj, 190, 447

\bibitem[{{Liebing} {et~al.}(2021){Liebing}, {Jeffers}, {Reiners}, \&
  {Zechmeister}}]{2021AA...654A.168L}
{Liebing}, F., {Jeffers}, S.~V., {Reiners}, A., \& {Zechmeister}, M. 2021,
  \aap, 654, A168

\bibitem[{{Lindegren} \& {Dravins}(2003)}]{2003AA...401.1185L}
{Lindegren}, L., \& {Dravins}, D. 2003, \aap, 401, 1185

\bibitem[{{Lindegren} {et~al.}(2012){Lindegren}, {Lammers}, {Hobbs},
  {O'Mullane}, {Bastian}, \& {Hern{\'a}ndez}}]{2012AA...538A..78L}
{Lindegren}, L., {Lammers}, U., {Hobbs}, D., {et~al.} 2012, \aap, 538, A78

\bibitem[{{Liu} {et~al.}(2021){Liu}, {Fang}, {Tian}, {Liu}, {Yang}, \&
  {Xue}}]{2021ApJS..254...20L}
{Liu}, J., {Fang}, M., {Tian}, H., {et~al.} 2021, \apjs, 254, 20

\bibitem[{{Liu} {et~al.}(2013){Liu}, {Magnier}, {Deacon}, {Allers}, {Dupuy},
  {Kotson}, {Aller}, {Burgett}, {Chambers}, {Draper}, {Hodapp}, {Jedicke},
  {Kaiser}, {Kudritzki}, {Metcalfe}, {Morgan}, {Price}, {Tonry}, \&
  {Wainscoat}}]{2013ApJ...777L..20L}
{Liu}, M.~C., {Magnier}, E.~A., {Deacon}, N.~R., {et~al.} 2013, \apjl, 777, L20

\bibitem[{{L{\"o}hner-B{\"o}ttcher} {et~al.}(2019){L{\"o}hner-B{\"o}ttcher},
  {Schmidt}, {Schlichenmaier}, {Steinmetz}, \&
  {Holzwarth}}]{2019AA...624A..57L}
{L{\"o}hner-B{\"o}ttcher}, J., {Schmidt}, W., {Schlichenmaier}, R.,
  {Steinmetz}, T., \& {Holzwarth}, R. 2019, \aap, 624, A57

\bibitem[{{Luhman} {et~al.}(2024){Luhman}, {Alves de Oliveira}, {Baraffe},
  {Chabrier}, {Geballe}, {Parker}, {Pendleton}, \&
  {Tremblin}}]{2024AJ....167...19L}
{Luhman}, K.~L., {Alves de Oliveira}, C., {Baraffe}, I., {et~al.} 2024, \aj,
  167, 19

\bibitem[{{Luri} {et~al.}(2018){Luri}, {Brown}, {Sarro}, {Arenou},
  {Bailer-Jones}, {Castro-Ginard}, {de Bruijne}, {Prusti}, {Babusiaux}, \&
  {Delgado}}]{2018AA...616A...9L}
{Luri}, X., {Brown}, A.~G.~A., {Sarro}, L.~M., {et~al.} 2018, \aap, 616, A9

\bibitem[{{Lutz} \& {Kelker}(1973)}]{1973PASP...85..573L}
{Lutz}, T.~E., \& {Kelker}, D.~H. 1973, \pasp, 85, 573

\bibitem[{{Lyra} \& {Porto de Mello}(2005)}]{2005AA...431..329L}
{Lyra}, W., \& {Porto de Mello}, G.~F. 2005, \aap, 431, 329

\bibitem[{{MacGregor} \& {Brenner}(1991)}]{1991ApJ...376..204M}
{MacGregor}, K.~B., \& {Brenner}, M. 1991, \apj, 376, 204

\bibitem[{{Madhusudhan} {et~al.}(2016){Madhusudhan}, {Ag{\'u}ndez}, {Moses}, \&
  {Hu}}]{2016SSRv..205..285M}
{Madhusudhan}, N., {Ag{\'u}ndez}, M., {Moses}, J.~I., \& {Hu}, Y. 2016, \ssr,
  205, 285

\bibitem[{{Malo} {et~al.}(2014){Malo}, {Doyon}, {Feiden}, {Albert},
  {Lafreni{\`e}re}, {Artigau}, {Gagn{\'e}}, \& {Riedel}}]{2014ApJ...792...37M}
{Malo}, L., {Doyon}, R., {Feiden}, G.~A., {et~al.} 2014, \apj, 792, 37

\bibitem[{{Malo} {et~al.}(2013){Malo}, {Doyon}, {Lafreni{\`e}re}, {Artigau},
  {Gagn{\'e}}, {Baron}, \& {Riedel}}]{2013ApJ...762...88M}
{Malo}, L., {Doyon}, R., {Lafreni{\`e}re}, D., {et~al.} 2013, \apj, 762, 88

\bibitem[{{Mamajek}(2009)}]{2009IAUS..258..375M}
{Mamajek}, E.~E. 2009, in The Ages of Stars, ed. E.~E. {Mamajek}, D.~R.
  {Soderblom}, \& R.~F.~G. {Wyse}, Vol. 258, 375--382

\bibitem[{{Mamajek} \& {Hillenbrand}(2008)}]{2008ApJ...687.1264M}
{Mamajek}, E.~E., \& {Hillenbrand}, L.~A. 2008, \apj, 687, 1264

\bibitem[{{Mann} {et~al.}(2019){Mann}, {Dupuy}, {Kraus}, {Gaidos}, {Ansdell},
  {Ireland}, {Rizzuto}, {Hung}, {Dittmann}, {Factor}, {Feiden}, {Martinez},
  {Ru{\'\i}z-Rodr{\'\i}guez}, \& {Thao}}]{2019ApJ...871...63M}
{Mann}, A.~W., {Dupuy}, T., {Kraus}, A.~L., {et~al.} 2019, \apj, 871, 63

\bibitem[{{Marois} {et~al.}(2008){Marois}, {Macintosh}, {Barman}, {Zuckerman},
  {Song}, {Patience}, {Lafreni{\`e}re}, \& {Doyon}}]{2008Sci...322.1348M}
{Marois}, C., {Macintosh}, B., {Barman}, T., {et~al.} 2008, Science, 322, 1348

\bibitem[{{McCarthy} \& {Wilhelm}(2014)}]{2014AJ....148...70M}
{McCarthy}, K., \& {Wilhelm}, R.~J. 2014, \aj, 148, 70

\bibitem[{{Meingast} \& {Alves}(2019)}]{2019AA...621L...3M}
{Meingast}, S., \& {Alves}, J. 2019, \aap, 621, L3

\bibitem[{{Meingast} {et~al.}(2019){Meingast}, {Alves}, \&
  {F{\"u}rnkranz}}]{2019AA...622L..13M}
{Meingast}, S., {Alves}, J., \& {F{\"u}rnkranz}, V. 2019, \aap, 622, L13

\bibitem[{{Meingast} {et~al.}(2021){Meingast}, {Alves}, \&
  {Rottensteiner}}]{2021AA...645A..84M}
{Meingast}, S., {Alves}, J., \& {Rottensteiner}, A. 2021, \aap, 645, A84

\bibitem[{{Mel'Nik} \& {Efremov}(1995)}]{1995AstL...21...10M}
{Mel'Nik}, A.~M., \& {Efremov}, Y.~N. 1995, Astronomy Letters, 21, 10

\bibitem[{{Mermilliod}(2000)}]{2000ASPC..198..105M}
{Mermilliod}, J.~C. 2000, in Astronomical Society of the Pacific Conference
  Series, Vol. 198, Stellar Clusters and Associations: Convection, Rotation,
  and Dynamos, ed. R.~{Pallavicini}, G.~{Micela}, \& S.~{Sciortino}, 105

\bibitem[{{Messina} {et~al.}(2010){Messina}, {Desidera}, {Turatto},
  {Lanzafame}, \& {Guinan}}]{2010AA...520A..15M}
{Messina}, S., {Desidera}, S., {Turatto}, M., {Lanzafame}, A.~C., \& {Guinan},
  E.~F. 2010, \aap, 520, A15

\bibitem[{{Mestel}(1968)}]{1968MNRAS.138..359M}
{Mestel}, L. 1968, \mnras, 138, 359

\bibitem[{{Meunier} {et~al.}(2017){Meunier}, {Lagrange}, {Mbemba Kabuiku},
  {Alex}, {Mignon}, \& {Borgniet}}]{2017AA...597A..52M}
{Meunier}, N., {Lagrange}, A.~M., {Mbemba Kabuiku}, L., {et~al.} 2017, \aap,
  597, A52

\bibitem[{{Michaud} \& {Charbonneau}(1991)}]{1991SSRv...57....1M}
{Michaud}, G., \& {Charbonneau}, P. 1991, \ssr, 57, 1

\bibitem[{{Miret-Roig} {et~al.}(2024){Miret-Roig}, {Alves}, {Barrado},
  {Burkert}, {Ratzenb{\"o}ck}, \& {Konietzka}}]{2024NatAs...8..216M}
{Miret-Roig}, N., {Alves}, J., {Barrado}, D., {et~al.} 2024, Nature Astronomy,
  8, 216

\bibitem[{{Miret-Roig} {et~al.}(2020){Miret-Roig}, {Galli}, {Brandner}, {Bouy},
  {Barrado}, {Olivares}, {Antoja}, {Romero-G{\'o}mez}, {Figueras}, \&
  {Lillo-Box}}]{2020AA...642A.179M}
{Miret-Roig}, N., {Galli}, P.~A.~B., {Brandner}, W., {et~al.} 2020, \aap, 642,
  A179

\bibitem[{{Moranta} {et~al.}(2022){Moranta}, {Gagn{\'e}}, {Couture}, \&
  {Faherty}}]{2022ApJ...939...94M}
{Moranta}, L., {Gagn{\'e}}, J., {Couture}, D., \& {Faherty}, J.~K. 2022, \apj,
  939, 94

\bibitem[{{Mu{\v{z}}i{\'c}} {et~al.}(2012){Mu{\v{z}}i{\'c}}, {Scholz}, {Geers},
  {Jayawardhana}, \& {Tamura}}]{2012ApJ...744..134M}
{Mu{\v{z}}i{\'c}}, K., {Scholz}, A., {Geers}, V., {Jayawardhana}, R., \&
  {Tamura}, M. 2012, \apj, 744, 134

\bibitem[{{Mu{\v{z}}i{\'c}} {et~al.}(2015){Mu{\v{z}}i{\'c}}, {Scholz}, {Geers},
  \& {Jayawardhana}}]{2015ApJ...810..159M}
{Mu{\v{z}}i{\'c}}, K., {Scholz}, A., {Geers}, V.~C., \& {Jayawardhana}, R.
  2015, \apj, 810, 159

\bibitem[{{Newton} {et~al.}(2022){Newton}, {Rampalli}, {Kraus}, {Mann},
  {Curtis}, {Vanderburg}, {Krolikowski}, {Huber}, {Petter}, {Bieryla},
  {Tofflemire}, {Thao}, {Wood}, {Kerr}, {Safanov}, {Strakhov}, {Ciardi},
  {Giacalone}, {Dressing}, {Gill}, {Savel}, {Collins}, {Brown}, {Murgas},
  {Isogai}, {Narita}, {Palle}, {Quinn}, {Eastman}, {F{\H{u}}r{\'e}sz}, {Shiao},
  {Daylan}, {Caldwell}, {Ricker}, {Vanderspek}, {Seager}, {Winn}, {Jenkins}, \&
  {Latham}}]{2022AJ....164..115N}
{Newton}, E.~R., {Rampalli}, R., {Kraus}, A.~L., {et~al.} 2022, \aj, 164, 115

\bibitem[{{Nordstroem} {et~al.}(1997){Nordstroem}, {Andersen}, \&
  {Andersen}}]{1997AA...322..460N}
{Nordstroem}, B., {Andersen}, J., \& {Andersen}, M.~I. 1997, \aap, 322, 460

\bibitem[{{Nordstr{\"o}m} {et~al.}(2004){Nordstr{\"o}m}, {Mayor}, {Andersen},
  {Holmberg}, {Pont}, {J{\o}rgensen}, {Olsen}, {Udry}, \&
  {Mowlavi}}]{2004AA...418..989N}
{Nordstr{\"o}m}, B., {Mayor}, M., {Andersen}, J., {et~al.} 2004, \aap, 418, 989

\bibitem[{{Odenkirchen} {et~al.}(2003){Odenkirchen}, {Grebel}, {Dehnen}, {Rix},
  {Yanny}, {Newberg}, {Rockosi}, {Mart{\'\i}nez-Delgado}, {Brinkmann}, \&
  {Pier}}]{2003AJ....126.2385O}
{Odenkirchen}, M., {Grebel}, E.~K., {Dehnen}, W., {et~al.} 2003, \aj, 126, 2385

\bibitem[{{Oh} {et~al.}(2017){Oh}, {Price-Whelan}, {Hogg}, {Morton}, \&
  {Spergel}}]{2017AJ....153..257O}
{Oh}, S., {Price-Whelan}, A.~M., {Hogg}, D.~W., {Morton}, T.~D., \& {Spergel},
  D.~N. 2017, \aj, 153, 257

\bibitem[{{Ortega} {et~al.}(2002){Ortega}, {de la Reza}, {Jilinski}, \&
  {Bazzanella}}]{2002ApJ...575L..75O}
{Ortega}, V.~G., {de la Reza}, R., {Jilinski}, E., \& {Bazzanella}, B. 2002,
  \apjl, 575, L75

\bibitem[{{Ortega} {et~al.}(2004){Ortega}, {de la Reza}, {Jilinski}, \&
  {Bazzanella}}]{2004ApJ...609..243O}
---. 2004, \apj, 609, 243

\bibitem[{{Oudmaijer} {et~al.}(1998){Oudmaijer}, {Groenewegen}, \&
  {Schrijver}}]{1998MNRAS.294L..41O}
{Oudmaijer}, R.~D., {Groenewegen}, M. A.~T., \& {Schrijver}, H. 1998, \mnras,
  294, L41

\bibitem[{{Parker}(1988)}]{1988ApJ...330..474P}
{Parker}, E.~N. 1988, \apj, 330, 474

\bibitem[{{Parker}(1993)}]{1993ApJ...408..707P}
---. 1993, \apj, 408, 707

\bibitem[{{Pasquini} {et~al.}(2011){Pasquini}, {Melo}, {Chavero}, {Dravins},
  {Ludwig}, {Bonifacio}, \& {de La Reza}}]{2011AA...526A.127P}
{Pasquini}, L., {Melo}, C., {Chavero}, C., {et~al.} 2011, \aap, 526, A127

\bibitem[{{Pecaut} \& {Mamajek}(2013)}]{2013ApJS..208....9P}
{Pecaut}, M.~J., \& {Mamajek}, E.~E. 2013, \apjs, 208, 9

\bibitem[{{Pecaut} \& {Mamajek}(2016)}]{2016MNRAS.461..794P}
---. 2016, \mnras, 461, 794

\bibitem[{{Popper}(1954)}]{1954ApJ...120..316P}
{Popper}, D.~M. 1954, \apj, 120, 316

\bibitem[{{Predehl} {et~al.}(2021){Predehl}, {Andritschke}, {Arefiev},
  {Babyshkin}, {Batanov}, {Becker}, {B{\"o}hringer}, {Bogomolov}, {Boller},
  {Borm}, {Bornemann}, {Br{\"a}uninger}, {Br{\"u}ggen}, {Brunner}, {Brusa},
  {Bulbul}, {Buntov}, {Burwitz}, {Burkert}, {Clerc}, {Churazov}, {Coutinho},
  {Dauser}, {Dennerl}, {Doroshenko}, {Eder}, {Emberger}, {Eraerds},
  {Finoguenov}, {Freyberg}, {Friedrich}, {Friedrich}, {F{\"u}rmetz},
  {Georgakakis}, {Gilfanov}, {Granato}, {Grossberger}, {Gueguen}, {Gureev},
  {Haberl}, {H{\"a}lker}, {Hartner}, {Hasinger}, {Huber}, {Ji}, {Kienlin},
  {Kink}, {Korotkov}, {Kreykenbohm}, {Lamer}, {Lomakin}, {Lapshov}, {Liu},
  {Maitra}, {Meidinger}, {Menz}, {Merloni}, {Mernik}, {Mican}, {Mohr},
  {M{\"u}ller}, {Nandra}, {Nazarov}, {Pacaud}, {Pavlinsky}, {Perinati},
  {Pfeffermann}, {Pietschner}, {Ramos-Ceja}, {Rau}, {Reiffers}, {Reiprich},
  {Robrade}, {Salvato}, {Sanders}, {Santangelo}, {Sasaki}, {Scheuerle},
  {Schmid}, {Schmitt}, {Schwope}, {Shirshakov}, {Steinmetz}, {Stewart},
  {Str{\"u}der}, {Sunyaev}, {Tenzer}, {Tiedemann}, {Tr{\"u}mper}, {Voron},
  {Weber}, {Wilms}, \& {Yaroshenko}}]{2021AA...647A...1P}
{Predehl}, P., {Andritschke}, R., {Arefiev}, V., {et~al.} 2021, \aap, 647, A1

\bibitem[{{Preibisch}(2012)}]{2012RAA....12....1P}
{Preibisch}, T. 2012, Research in Astronomy and Astrophysics, 12, 1

\bibitem[{{Rebolo}(1991)}]{1991IAUS..145...85R}
{Rebolo}, R. 1991, in Evolution of Stars: the Photospheric Abundance
  Connection, ed. G.~{Michaud} \& A.~V. {Tutukov}, Vol. 145, 85

\bibitem[{{Rebolo} {et~al.}(1996){Rebolo}, {Martin}, {Basri}, {Marcy}, \&
  {Zapatero-Osorio}}]{1996ApJ...469L..53R}
{Rebolo}, R., {Martin}, E.~L., {Basri}, G., {Marcy}, G.~W., \&
  {Zapatero-Osorio}, M.~R. 1996, \apjl, 469, L53

\bibitem[{{Ricker} {et~al.}(2015){Ricker}, {Winn}, {Vanderspek}, {Latham},
  {Bakos}, {Bean}, {Berta-Thompson}, {Brown}, {Buchhave}, {Butler}, {Butler},
  {Chaplin}, {Charbonneau}, {Christensen-Dalsgaard}, {Clampin}, {Deming},
  {Doty}, {De Lee}, {Dressing}, {Dunham}, {Endl}, {Fressin}, {Ge}, {Henning},
  {Holman}, {Howard}, {Ida}, {Jenkins}, {Jernigan}, {Johnson}, {Kaltenegger},
  {Kawai}, {Kjeldsen}, {Laughlin}, {Levine}, {Lin}, {Lissauer}, {MacQueen},
  {Marcy}, {McCullough}, {Morton}, {Narita}, {Paegert}, {Palle}, {Pepe},
  {Pepper}, {Quirrenbach}, {Rinehart}, {Sasselov}, {Sato}, {Seager},
  {Sozzetti}, {Stassun}, {Sullivan}, {Szentgyorgyi}, {Torres}, {Udry}, \&
  {Villasenor}}]{2015JATIS...1a4003R}
{Ricker}, G.~R., {Winn}, J.~N., {Vanderspek}, R., {et~al.} 2015, Journal of
  Astronomical Telescopes, Instruments, and Systems, 1, 014003

\bibitem[{{Rogers} {et~al.}(2011){Rogers}, {Bodenheimer}, {Lissauer}, \&
  {Seager}}]{2011ApJ...738...59R}
{Rogers}, L.~A., {Bodenheimer}, P., {Lissauer}, J.~J., \& {Seager}, S. 2011,
  \apj, 738, 59

\bibitem[{{R{\"o}ser} \& {Schilbach}(2019)}]{2019AA...627A...4R}
{R{\"o}ser}, S., \& {Schilbach}, E. 2019, \aap, 627, A4

\bibitem[{{R{\"o}ser} {et~al.}(2019){R{\"o}ser}, {Schilbach}, \&
  {Goldman}}]{2019AA...621L...2R}
{R{\"o}ser}, S., {Schilbach}, E., \& {Goldman}, B. 2019, \aap, 621, L2

\bibitem[{{Safsten} {et~al.}(2020){Safsten}, {Dawson}, \&
  {Wolfgang}}]{2020DDA....5110103S}
{Safsten}, E., {Dawson}, R., \& {Wolfgang}, A. 2020, in AAS/Division of
  Dynamical Astronomy Meeting, Vol.~52, AAS/Division of Dynamical Astronomy
  Meeting, 101.03

\bibitem[{{Schneider} {et~al.}(2023){Schneider}, {Burgasser}, {Bruursema},
  {Munn}, {Vrba}, {Caselden}, {Kabatnik}, {Rothermich}, {Sainio}, {Bickle},
  {Dahm}, {Meisner}, {Kirkpatrick}, {Su{\'a}rez}, {Gagn{\'e}}, {Faherty},
  {Vos}, {Kuchner}, {Williams}, {Bardalez Gagliuffi}, {Aganze}, {Hsu},
  {Theissen}, {Cushing}, {Marocco}, {Casewell}, \& {Backyard Worlds: Planet 9
  Collaboration}}]{2023ApJ...943L..16S}
{Schneider}, A.~C., {Burgasser}, A.~J., {Bruursema}, J., {et~al.} 2023, \apjl,
  943, L16

\bibitem[{{Scholz} {et~al.}(2012){Scholz}, {Jayawardhana}, {Muzic}, {Geers},
  {Tamura}, \& {Tanaka}}]{2012ApJ...756...24S}
{Scholz}, A., {Jayawardhana}, R., {Muzic}, K., {et~al.} 2012, \apj, 756, 24

\bibitem[{{Sim} {et~al.}(2019){Sim}, {Lee}, {Ann}, \&
  {Kim}}]{2019JKAS...52..145S}
{Sim}, G., {Lee}, S.~H., {Ann}, H.~B., \& {Kim}, S. 2019, Journal of Korean
  Astronomical Society, 52, 145

\bibitem[{{Skuljan} {et~al.}(1999){Skuljan}, {Hearnshaw}, \&
  {Cottrell}}]{1999MNRAS.308..731S}
{Skuljan}, J., {Hearnshaw}, J.~B., \& {Cottrell}, P.~L. 1999, \mnras, 308, 731

\bibitem[{{Skumanich}(1972)}]{1972ApJ...171..565S}
{Skumanich}, A. 1972, \apj, 171, 565

\bibitem[{{Soderblom}(2010)}]{2010ARAA..48..581S}
{Soderblom}, D.~R. 2010, \araa, 48, 581

\bibitem[{{Soderblom} {et~al.}(1991){Soderblom}, {Duncan}, \&
  {Johnson}}]{1991ApJ...375..722S}
{Soderblom}, D.~R., {Duncan}, D.~K., \& {Johnson}, D. R.~H. 1991, \apj, 375,
  722

\bibitem[{{Soderblom} {et~al.}(2014){Soderblom}, {Hillenbrand}, {Jeffries},
  {Mamajek}, \& {Naylor}}]{2014prpl.conf..219S}
{Soderblom}, D.~R., {Hillenbrand}, L.~A., {Jeffries}, R.~D., {Mamajek}, E.~E.,
  \& {Naylor}, T. 2014, in Protostars and Planets VI, ed. H.~{Beuther}, R.~S.
  {Klessen}, C.~P. {Dullemond}, \& T.~{Henning}, 219--241

\bibitem[{{Soderblom} {et~al.}(1993){Soderblom}, {Jones}, {Balachandran},
  {Stauffer}, {Duncan}, {Fedele}, \& {Hudon}}]{1993AJ....106.1059S}
{Soderblom}, D.~R., {Jones}, B.~F., {Balachandran}, S., {et~al.} 1993, \aj,
  106, 1059

\bibitem[{{Song} {et~al.}(2003){Song}, {Zuckerman}, \&
  {Bessell}}]{2003ApJ...599..342S}
{Song}, I., {Zuckerman}, B., \& {Bessell}, M.~S. 2003, \apj, 599, 342

\bibitem[{{Steenbeck} \& {Krause}(1969)}]{1969AN....291...49S}
{Steenbeck}, M., \& {Krause}, F. 1969, Astronomische Nachrichten, 291, 49

\bibitem[{{Suzuki} {et~al.}(2013){Suzuki}, {Imada}, {Kataoka}, {Kato},
  {Matsumoto}, {Miyahara}, \& {Tsuneta}}]{2013PASJ...65...98S}
{Suzuki}, T.~K., {Imada}, S., {Kataoka}, R., {et~al.} 2013, \pasj, 65, 98

\bibitem[{{Tang} {et~al.}(2019){Tang}, {Pang}, {Yuan}, {Chen}, {Hong},
  {Goldman}, {Just}, {Shukirgaliyev}, \& {Lin}}]{2019ApJ...877...12T}
{Tang}, S.-Y., {Pang}, X., {Yuan}, Z., {et~al.} 2019, \apj, 877, 12

\bibitem[{{Tofflemire} {et~al.}(2021){Tofflemire}, {Rizzuto}, {Newton},
  {Kraus}, {Mann}, {Vanderburg}, {Nelson}, {Hawkins}, {Wood}, {Zhou}, {Quinn},
  {Howell}, {Collins}, {Schwarz}, {Stassun}, {Bouma}, {Essack}, {Osborn},
  {Boyd}, {F{\H{u}}r{\'e}sz}, {Glidden}, {Twicken}, {Wohler}, {McLean},
  {Ricker}, {Vanderspek}, {Latham}, {Seager}, {Winn}, \&
  {Jenkins}}]{2021AJ....161..171T}
{Tofflemire}, B.~M., {Rizzuto}, A.~C., {Newton}, E.~R., {et~al.} 2021, \aj,
  161, 171

\bibitem[{{Torres} {et~al.}(2008){Torres}, {Quast}, {Melo}, \&
  {Sterzik}}]{2008hsf2.book..757T}
{Torres}, C.~A.~O., {Quast}, G.~R., {Melo}, C.~H.~F., \& {Sterzik}, M.~F. 2008,
  in Handbook of Star Forming Regions, Volume II, ed. B.~{Reipurth}, Vol.~5,
  757

\bibitem[{{Trifonov} {et~al.}(2020){Trifonov}, {Tal-Or}, {Zechmeister},
  {Kaminski}, {Zucker}, \& {Mazeh}}]{2020AA...636A..74T}
{Trifonov}, T., {Tal-Or}, L., {Zechmeister}, M., {et~al.} 2020, \aap, 636, A74

\bibitem[{{Truemper}(1982)}]{1982AdSpR...2d.241T}
{Truemper}, J. 1982, Advances in Space Research, 2, 241

\bibitem[{{Unterborn} {et~al.}(2022){Unterborn}, {Foley}, {Desch}, {Young},
  {Vance}, {Chiffelle}, \& {Kane}}]{2022ApJ...930L...6U}
{Unterborn}, C.~T., {Foley}, B.~J., {Desch}, S.~J., {et~al.} 2022, \apjl, 930,
  L6

\bibitem[{{van Leeuwen}(2007)}]{2007AA...474..653V}
{van Leeuwen}, F. 2007, \aap, 474, 653

\bibitem[{{Veras} \& {Tout}(2012)}]{2012MNRAS.422.1648V}
{Veras}, D., \& {Tout}, C.~A. 2012, \mnras, 422, 1648

\bibitem[{{Voges} {et~al.}(1999){Voges}, {Aschenbach}, {Boller},
  {Br{\"a}uninger}, {Briel}, {Burkert}, {Dennerl}, {Englhauser}, {Gruber},
  {Haberl}, {Hartner}, {Hasinger}, {K{\"u}rster}, {Pfeffermann}, {Pietsch},
  {Predehl}, {Rosso}, {Schmitt}, {Tr{\"u}mper}, \&
  {Zimmermann}}]{1999AA...349..389V}
{Voges}, W., {Aschenbach}, B., {Boller}, T., {et~al.} 1999, \aap, 349, 389

\bibitem[{{Vos} {et~al.}(2023){Vos}, {Burningham}, {Faherty}, {Alejandro},
  {Gonzales}, {Calamari}, {Bardalez Gagliuffi}, {Visscher}, {Tan}, {Morley},
  {Marley}, {Gemma}, {Whiteford}, {Gaarn}, \& {Park}}]{2023ApJ...944..138V}
{Vos}, J.~M., {Burningham}, B., {Faherty}, J.~K., {et~al.} 2023, \apj, 944, 138

\bibitem[{{{\v{Z}}erjal} {et~al.}(2023){{\v{Z}}erjal}, {Ireland}, {Crundall},
  {Krumholz}, \& {Rains}}]{2023MNRAS.519.3992Z}
{{\v{Z}}erjal}, M., {Ireland}, M.~J., {Crundall}, T.~D., {Krumholz}, M.~R., \&
  {Rains}, A.~D. 2023, \mnras, 519, 3992

\bibitem[{{Weber} \& {Davis}(1967)}]{1967ApJ...148..217W}
{Weber}, E.~J., \& {Davis}, Leverett, J. 1967, \apj, 148, 217

\bibitem[{{Wegner}(1979)}]{1979AJ.....84..650W}
{Wegner}, G. 1979, \aj, 84, 650

\bibitem[{{Wilking} \& {Lada}(1983)}]{1983ApJ...274..698W}
{Wilking}, B.~A., \& {Lada}, C.~J. 1983, \apj, 274, 698

\bibitem[{{Wright} \& {Eastman}(2014)}]{2014PASP..126..838W}
{Wright}, J.~T., \& {Eastman}, J.~D. 2014, \pasp, 126, 838

\bibitem[{{Zapatero Osorio} {et~al.}(2006){Zapatero Osorio}, {Mart{\'\i}n},
  {Bouy}, {Tata}, {Deshpande}, \& {Wainscoat}}]{2006ApJ...647.1405Z}
{Zapatero Osorio}, M.~R., {Mart{\'\i}n}, E.~L., {Bouy}, H., {et~al.} 2006,
  \apj, 647, 1405

\bibitem[{{Zhang} {et~al.}(2021){Zhang}, {Liu}, {Best}, {Dupuy}, \&
  {Siverd}}]{2021ApJ...911....7Z}
{Zhang}, Z., {Liu}, M.~C., {Best}, W. M.~J., {Dupuy}, T.~J., \& {Siverd}, R.~J.
  2021, \apj, 911, 7

\bibitem[{{Zuckerman} \& {Song}(2004)}]{2004ARAA..42..685Z}
{Zuckerman}, B., \& {Song}, I. 2004, \araa, 42, 685

\bibitem[{{Zuckerman} \& {Song}(2009)}]{2009AA...493.1149Z}
---. 2009, \aap, 493, 1149

\bibitem[{{Zuckerman} {et~al.}(2004){Zuckerman}, {Song}, \&
  {Bessell}}]{2004ApJ...613L..65Z}
{Zuckerman}, B., {Song}, I., \& {Bessell}, M.~S. 2004, \apjl, 613, L65

\bibitem[{{Zuckerman} {et~al.}(2001){Zuckerman}, {Song}, {Bessell}, \&
  {Webb}}]{2001ApJ...562L..87Z}
{Zuckerman}, B., {Song}, I., {Bessell}, M.~S., \& {Webb}, R.~A. 2001, \apjl,
  562, L87

\bibitem[{{Zuckerman} \& {Webb}(2000)}]{2000ApJ...535..959Z}
{Zuckerman}, B., \& {Webb}, R.~A. 2000, \apj, 535, 959

\end{thebibliography}

\end{document}